\documentclass[fleqn,12pt]{olplainarticle}
\usepackage{lmodern} % improved font
\usepackage{changepage}
\usepackage{setspace}
\usepackage{hyperref}
\usepackage{pdfpages}
\usepackage{wrapfig}
\usepackage{acronym}
\usepackage{tikz}
\usepackage{cleveref}
\crefname{figure}{figure}{figures}
\usepackage{textcomp}
\usepackage{textgreek}
\usepackage{enumitem}\usepackage{multirow}
\usepackage{subcaption}
\title{Electrical Control over Volatile Mott Switching through Non-Volatile Memory Effects}

\author[1]{Amihai Kronman\textsuperscript{\textdagger,}}
\author[1]{Gil Levi\textsuperscript{\textdagger,}}
\author[1]{Yoav Kalcheim}
\affil[1]{\textit{Department of Materials Science and Engineering, Technion - Israel Institute of Technology}, Haifa, Israel}


\begin{abstract}
Vanadium Dioxide (VO\textsubscript{2}) and Vanadium Sesquioxide (V\textsubscript{2}O\textsubscript{3}) are Mott insulators that undergo an Insulator-to-Metal Transition (IMT) at $\sim$340K and $\sim$160K, respectively, manifested as an orders-of-magnitude reduction in their electrical resistance. These transitions provide the physical basis for their Volatile Resistive Switching (VRS) behavior, making them promising candidates for threshold-switching devices. Here, we show that the volatile switching in these materials can be strongly affected by non-volatile processes associated with the creation and annihilation of the conducting filament. These processes give rise to pronounced memory effects in which the initial switching voltage substantially exceeds that of subsequent cycles. We identify two distinct mechanisms underlying this behavior in different thermal regimes. In the phase-coexistence regime, a memory effect is observed in V\textsubscript{2}O\textsubscript{3}, arising from the spatial redistribution of metallic and insulating domains. Well below the hysteresis regime, an additional electroforming-like memory effect is observed in both VO\textsubscript{2} and V\textsubscript{2}O\textsubscript{3} which is attributed to the formation and/or migration of defects under the influence of a high electric field and a current surge associated with the IMT-driven switching. Using a protective internal resistor and a multi-step writing protocol, this normally destructive process can be harnessed to tune the switching voltage and power over a wide temperature range. These results demonstrate a route toward controlled programming of switching parameters in IMT-based resistive switching devices. 

\end{abstract}
 
\begin{document}

\flushbottom
\maketitle

\begingroup
\renewcommand{\thefootnote}{\ensuremath{\dagger}}
\footnotetext{A.K. and G.L. contributed equally to this work.}
\endgroup

\thispagestyle{empty}

\section*{Introduction}

Resistive Switching (RS) is a phenomenon where the electrical resistance of a device can be modified from a High-Resistance-State (HRS) to a Low-Resistance-State (LRS) and vice versa by applying an external electric stimulus (current/voltage). The RS can be either volatile, where the system reverts to its initial state upon the cessation of the external signal (also known as ‘threshold switching’), or non-volatile, where the transition from HRS to LRS, or vice versa, persists after the applied electrical stimulus is removed \cite{del_valle_challenges_2018}. VRS may originate from several distinct physical mechanisms, including ion migration \cite{banerjee_deep_2021}, redox-induced filament formation and rupture \cite{sun_design_2020, wang_realizing_2019}, ovonic switching \cite{ho_lee_threshold_2012,koo_zn1xtex_2018}, or electronic phase transitions \cite{schneble_electrically-driven_2024,janod_resistive_2015,valmianski_origin_2018, pickett_sub-100_2012}. These devices are critical components in emerging technologies such as Neuromorphic Computing \cite{zhao_mott_2025,qiu_reconfigurable_2024,islam_electro-optical_2025,stoliar_leakyintegrateandfire_2017, maher_highly_2024,kim_mott_2024}, oscillators\cite{li_computational_2024, kim_electrical_2010,ma_mott_2020}, and as selectors in crossbar memory arrays \cite{ban_advances_2025,jeonghwan_song_threshold_2015,wang_realizing_2019, cha_comprehensive_2016,son_excellent_2011}. Consequently, achieving precise control over operational parameters, most notably the switching voltage and power, is vital for the development of scalable and energy-efficient VRS applications.\vspace{6pt}

In this work, we focus on VRS driven by an electrically induced insulator-to-metal transition (IMT) in VO\textsubscript{2} and V\textsubscript{2}O\textsubscript{3}. These vanadium oxides are prototypical correlated electron materials that exhibit an IMT accompanied by a several orders-of-magnitude changes in resistivity \cite{imada_metal-insulator_1998}, making them ideal materials for IMT-based resistive switching devices. In V\textsubscript{2}O\textsubscript{3} the transition involves a change in resistivity of up to 6 orders-of-magnitude, coupled with a structural and magnetic transformation \cite{morin_oxides_1959}.  Below the critical temperature ($\sim$160 K), V\textsubscript{2}O\textsubscript{3} is in an insulating antiferromagnetic monoclinic phase, while above this temperature it transforms into a metallic paramagnetic rhombohedral phase \cite{dernier_crystal_1970, huang_magnetic_2025}. In VO\textsubscript{2} the IMT is characterized by up to 4 orders of magnitude change in resistivity, coupled with a structural transition from a low-temperature insulating monoclinic phase to a high-temperature metallic rutile phase \cite{zylbersztejn_metal-insulator_1975,shabalin_nanoscale_2020, liu_recent_2018}.  The critical transition temperature of VO\textsubscript{2} is $\sim$340K, well-suited for operation in ambient conditions.\vspace{6pt}

When incorporated into thin-film devices, both oxides demonstrate reproducible VRS behavior linked to the electrically induced IMT \cite{guenon_electrical_2013,radu_switching_2015}. This behavior originates from the current-induced formation of a metallic filamentary path, which subsequently reverts to the insulating state upon the cessation of the electrical stimulus \cite{duchene_filamentary_1971,del_valle_subthreshold_2019,lange_imaging_2021, kumar_local_2013}.\vspace{6pt}

Although the formation and annihilation of the conducting filament is repeatable, its switching voltage and power, can vary between cycles. When such variations depend on the previous electrical or thermal history of the device, they are referred to here as memory effects.\vspace{6pt}

One type of memory effect arises from the hysteretic nature of the IMT in V\textsubscript{2}O\textsubscript{3} and VO\textsubscript{2}. In these materials, the transition from the insulating to the metallic state and the reverse transition occur at different critical temperatures during heating and cooling. Between these critical temperatures, there is a phase coexistence regime of insulating and metallic domains. In the context of IMT-based switching, this hysteresis produces a history-dependent switching voltage: the critical voltage required to switch from HRS to LRS depends on the thermal history of the device, for example whether switching is performed on the cooling branch or the heating branch of the transition \cite{rana_resistive_2020}.\vspace{6pt}

A second type of memory effect commonly observed in RS devices, both volatile \cite{conti_electroforming_2026, nandi_understanding_2020} and non-volatile \cite{ielmini_electroforming_2016}, is electroforming. In this process, the voltage required for switching is higher during the first IV sweep than in subsequent sweeps. In non-volatile RS devices, the role of electroforming is relatively straightforward: the first sweep creates a conducting filament, whereas from the second sweep onward, only a small region of the filament is ruptured and reconnected \cite{buckwell_conductance_2015, wei_threedimensional_2023}. In IMT-based devices, electroforming was also observed, where it introduced substantial structural modifications to the film, including amorphization and the formation of additional oxide phases \cite{conti_electroforming_2026, del_valle_electrically_2017}. The violent and stochastic nature of electroforming poses a significant challenge for the development of reliable VRS devices, where reproducible and controllable switching parameters are essential.

These considerations raise several key questions. What are the mechanisms responsible for memory effects in vanadium oxide VRS devices? What is the interplay between IMT-based switching dynamics and electroforming? And can this interplay be harnessed to achieve controlled modulation of switching parameters?
\vspace{6pt}

In this work, we systematically investigate the nature of memory effects observed in two vanadium oxides: V\textsubscript{2}O\textsubscript{3} and VO\textsubscript{2}. We identify two distinct types of memory effects, which occur in different temperature regimes and originate from different physical mechanisms. Within the phase-coexistence regime, V\textsubscript{2}O\textsubscript{3} exhibits a memory effect that is distinct from the conventional memory effect associated with the heating and cooling branches of the hysteresis loop. Even when the device remains on the cooling branch throughout the measurement, the first IV sweep differs systematically from subsequent sweeps. This behavior suggests that the initial switching event redistributes the metallic and insulating domains in a manner which is more conducive to filament formation than the initial domain configuration. The effect is reversible and can be erased by thermal cycling (i.e. heating the device to the fully metallic state and cooling back). In contrast, no analogous memory effect is observed in VO\textsubscript{2}.\vspace{6pt}

At lower temperatures, in both materials, we identify an intermediate regime in which the IV sweeps are highly repeatable, and no clear memory effect is observed. Upon further cooling in the fully insulating state, both V\textsubscript{2}O\textsubscript{3} and VO\textsubscript{2} exhibit a similar memory effect where the initial switching voltage is higher than in subsequent switching events. However, unlike the phase-coexistence memory, this memory effect is not erased by thermal cycling. We attribute this low-temperature memory effect to mild electroforming, where defect formation and/or migration is induced by the combination of a high electric field and current surge due to the initial switching event.\vspace{6pt}

Finally, we show the practical importance of the low temperature memory effect for device operation. By incorporating an internal resistor that limits the current surge, electroforming can be performed gradually at low temperatures, allowing defects to be introduced in a controlled manner rather than through an abrupt, aggressive forming event which damages the device. This approach enables tuning of the switching voltage over a wide temperature range, including the intermediate regime when no memory effect is observed, providing a pathway for enhanced all-electrical control of the VRS parameters, such as switching voltage and power.

\section*{Methods}
\subsection{Thin-film Growth}
Thin films of V\textsubscript{2}O\textsubscript{3} with thickness of $\sim$100nm were grown on A-cut Sapphire by sputtering from a 1.5" V\textsubscript{2}O\textsubscript{3} target in an Argon environment with substrate temperature of $\sim$650°C and pressure of 5mTorr. Thin films of VO\textsubscript{2} with thickness of $\sim$100nm were grown on A-cut Sapphire by reactive sputtering from a 1.5" V\textsubscript{2}O\textsubscript{3} target in a 95\% Argon 5\% Oxygen environment with substrate temperature of $\sim$400°C and pressure of 3mTorr.

\subsection{Lithography of devices}
After deposition, a grid of devices was fabricated on the films using a standard photolithography procces with e-beam evaporation of Ti (15nm)/Au (75nm) electrodes and a lift-off process. Each device is composed of a small gap of $\sim$4µm between the electrodes, and two larger gaps in the middle of each electrode (Figure \ref{fig: RT of films} inset). The larger gaps act as internal resistors that suppress current surges following the IMT, thereby protecting the device. 

\subsection{Electrical transport measurements}
Resistance vs. Temperature (R(T)) Measurements and Current vs. Voltage (IV) measurements were performed using the Lakeshore TTPX Cryogenic probe station. Temperature control was provided by the Lakeshore 336 Temperature Controller, the Keithley 2450 Source Meter Unit (SMU) was used for sourcing and measuring electric signals. 
The procedure for performing an IV measurement with the internal resistor was as follows: First, an IV measurement of the internal resistor only was performed, ensuring that it did not switch under the measurement's conditions. Afterwards, an IV measurement was performed on the internal resistor and the gap in series, and then the voltage on the internal resistor was subtracted from the total voltage to extract the IV curve of the gap.

\subsection{Scanning Electron Microscopy (SEM) imaging}
SEM imaging was performed using the SEM mode of the FEI Helios 5 PFIB DualBeam at an accelerating voltage of 5kV. Images were acquired using a secondary electron detector at a working distance of 3.9mm.

\subsection{Raman measurements}
Raman measurements were performed using a Witec Alpha300 Raman microscope equipped with a 532nm laser. The laser power was set to 3mW, and the spectra were acquired using a 1200l/mm grating. Each spectrum was obtained by averaging 30 measurements, with an integration time of 1s per measurement.

\section*{Results}
The as-grown vanadium oxide films exhibit characteristic hysteretic R(T) curves, with transition temperatures of $\sim$150K for V\textsubscript{2}O\textsubscript{3} and $\sim$340K for VO\textsubscript{2} (Fig. 1). Complementary X-ray diffraction measurements indicate the growth of single-phase, highly oriented films with out-of-plane orientation of (110) in V\textsubscript{2}O\textsubscript{3} and (200) and (002) in VO\textsubscript{2} (Supplementary figure S1).

\begin{figure}[!htbp]
\centering
\includegraphics[width=0.7\textwidth]{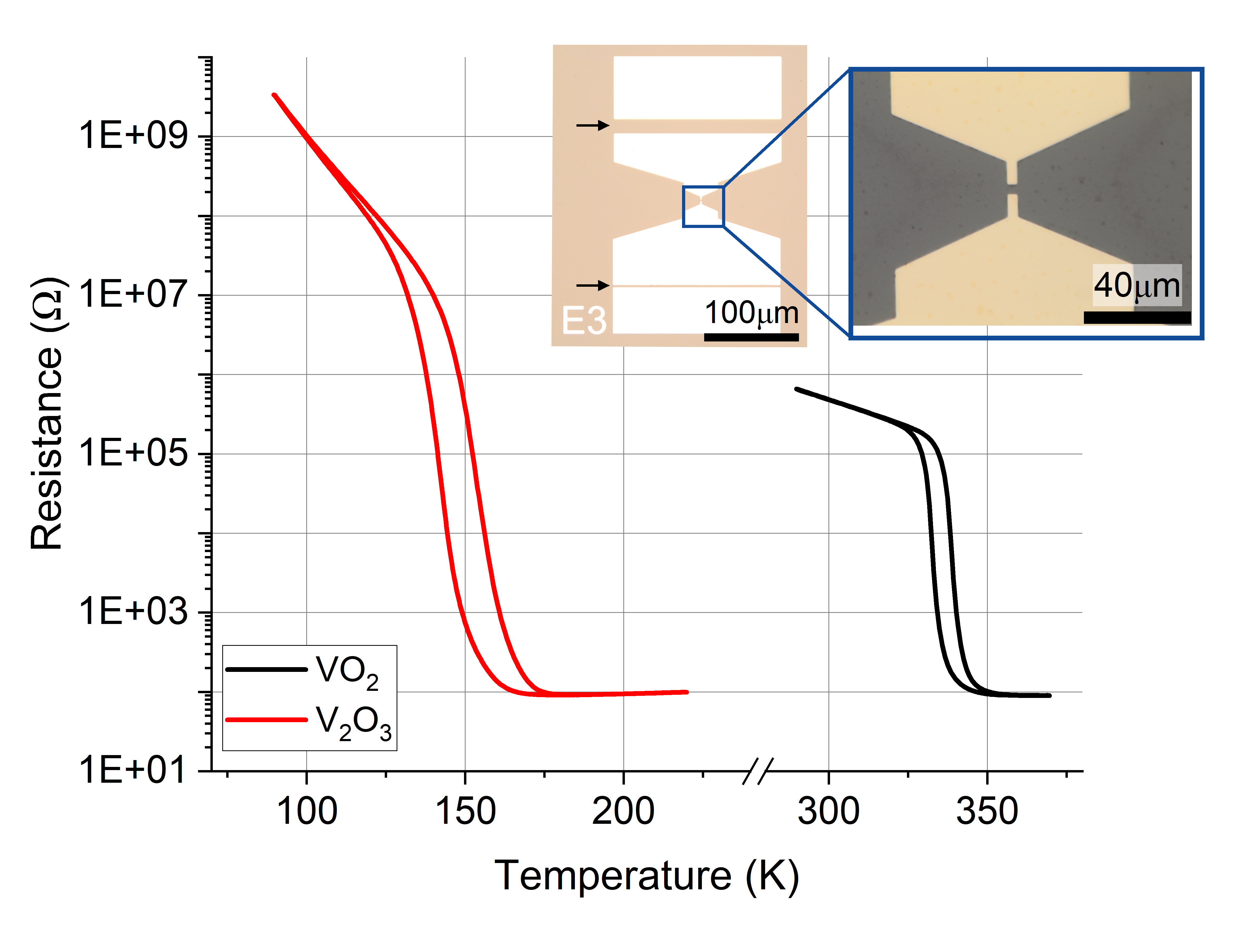}
\caption{RT curves of the as-grown V\textsubscript{2}O\textsubscript{3} and VO\textsubscript{2} thin films. Inset: Optical micrograph of a fabricated device. The black arrows indicate the location of internal resistors, the upper resistor was used in this work.}
\label{fig: RT of films}
\end{figure}

In a standard two-terminal configuration, performing Current vs. Voltage (IV) measurements on devices often leads to irreversible damage upon reaching the critical switching voltage (V\textsubscript{sw}). Because the formation of the conducting filament occurs on a nanosecond-or-less timescale \cite{brockman_subnanosecond_2014,zhou_voltage-triggered_2013, mayer_tunneling_2015}, far less than the response time of the current source, the voltage remains momentarily pinned at V\textsubscript{sw} while the device resistance drops sharply. This triggers a large current surge that damages the device (Supplementary figure S2). The inclusion of a resistor in series in the device's circuit can in principle suppress this current surge, acting as a current-limiter after switching until the current source readjusts to the drop in resistance. However, there is an additional source for a power surge on the circuit which is the fast discharge of parasitic capacitance in the circuit through the newly formed metallic filament. When using an external resistor, the parasitic capacitance of the wiring between the resistor and the device is large enough to cause permanent damage to the device as well (See supplementary section S3).\vspace{6pt}

\begin{figure}[!htbp]
     \centering
     \includegraphics[width=\linewidth]{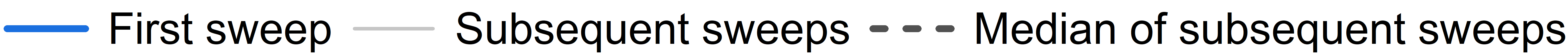}
     \begin{subfigure}[h]{0.48\textwidth}
         \centering
         \includegraphics[width=\linewidth]{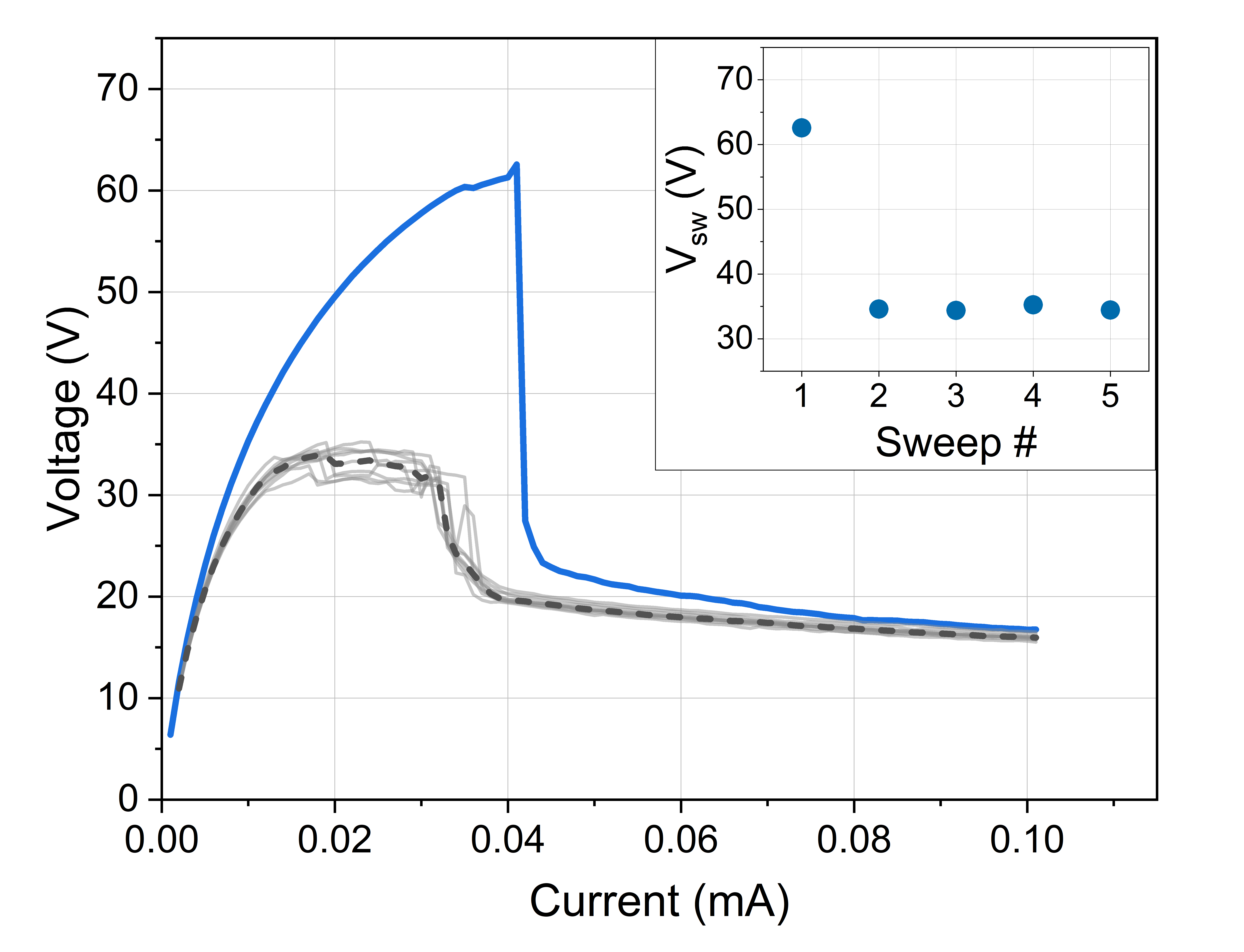}
         \caption{}
         \label{fig: Low T memory V2O3}
     \end{subfigure}
     \hfill
     \begin{subfigure}[h]{0.48\linewidth}
         \centering
         \includegraphics[width=\textwidth]{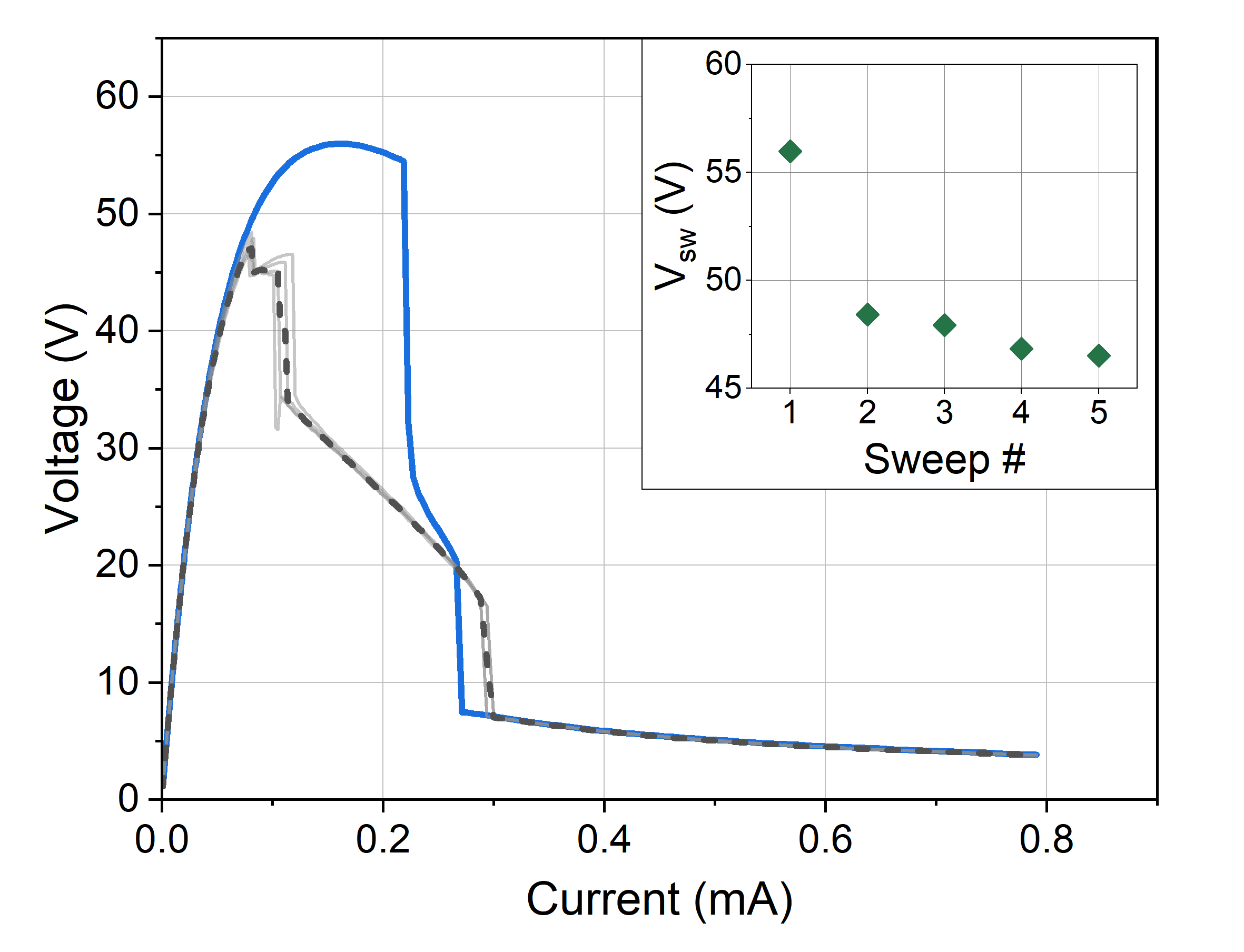}
         \caption{}
         \label{fig: Low T memory vo2}
     \end{subfigure}
     \hfill
    \caption{Memory effect in \textbf{(a)} V\textsubscript{2}O\textsubscript{3} (T=110K) and \textbf{(b)} VO\textsubscript{2} (T=250K) devices after incorporating an internal resistor. The inset shows the extracted switching voltage from each IV sweep.}
    \label{fig:low T memory effect}
\end{figure}

To circumvent this, we incorporated an internal series resistor into the device geometry (as indicated by the arrows  in the inset of Figure \ref{fig: RT of films}) which consists of a significantly larger gap of the same oxide (V\textsubscript{2}O\textsubscript{3} and VO\textsubscript{2}). Due to its larger electrode separation and width, this resistor possesses a higher V\textsubscript{sw} and retains a nearly constant high resistance state while the narrower gap switches (supplementary figure S3.1). The proximity of the internal resistor to the gap greatly reduces the amount of parasitic capacitance, which strongly limits the current which flows through the metallic filament after its creation, preventing surge-induced destruction (See supplementary section S3). Additionally, the Internal resistor maintains a nearly constant series-resistor to gap resistance ratio as the IV measurement temperature is varied, preserving comparable current-limiting conditions across the investigated temperature range. This device architecture therefore enables stable, consecutive IV cycles over a broad temperature range without adjustment of the resistor value during the measurement. This stability is maintained during fully insulating state, where switching would otherwise become destructive. Leveraging the protection provided by the internal resistor, multiple consecutive IV sweeps were performed at low temperatures. Despite the reduction of the power surge, it is not fully eliminated, and the IV curves revealed a significant decrease in the switching voltage between the initial IV sweep and subsequent measurements (Figure \ref{fig:low T memory effect}). The figures presented in this work show only the increasing-current branch of each IV sweep. 

Notably, this substantial decrease in switching voltage is accompanied by at most a marginal reduction in the insulating phase resistance (See Supplementary section S4), without any visible damage to the device or a significant change to the shape of the R(T) curve, unlike the significant changes observed for conventional electroforming in VO\textsubscript{x} devices \cite{conti_electroforming_2026, del_valle_electrically_2017}. Such a mild difference between the device’s state before and after the first IV sweep suggests that the system relaxed into a kinetically favorable metastable state following the first IV sweep, though the specific nature of this state is not clear. \vspace{6pt}

We propose three hypotheses for the mechanisms behind the observed memory effects:\vspace{6pt}

In the first hypothesis, the "Metallic-domain redistribution" hypothesis, the memory effect originates from a change in the spatial distribution of metallic domains following the first IV sweep. Within the phase-coexistence regime, metallic and insulating domains coexist thermodynamically, and the creation and annihilation of the conducting filament may redistribute the pre-existing metallic domains. Although the resulting metallic domains may not form a percolating path and therefore have only a limited effect on the total device resistance, their altered spatial configuration can lower the critical voltage required to reestablish a conducting filament in subsequent IV sweeps.

A related mechanism may also occur below the hysteresis regime, where previous studies have reported isolated metallic domains that persist as metastable remnants after the system is cooled into the insulating phase \cite{ramirez_first-order_2009, del_valle_subthreshold_2019}. In this case, those kinetically-hindered metallic domains are concentrated in the filament path following its annihilation and facilitate the formation of the filament at a lower switching voltage in subsequent switching events. As temperature is reduced further below the hysteresis regime, this memory effect is expected to diminish because the metallic domains become less stable\cite{del_valle_subthreshold_2019}. In both cases, thermal cycling is expected to restore the original domain configuration and recover the initial high switching voltage. \vspace{6pt}

In the second hypothesis, the "defects" hypothesis, the memory effect originates from the current surge that follows the rapid formation of the metallic filament. This surge, accompanied by the high electric field applied on the device during the IV measurement, induces the formation and/or migration of defects along the current path. Defects have been reported to reduce the critical switching voltage in vanadium oxide devices \cite{kalcheim_non-thermal_2020,wickramaratne_role_2019,ghazikhanian_resistive_2023}, and therefore may lower the voltage required to re-create the conducting filament in subsequent IV sweeps. Although this defect concentration is insufficient to significantly reduce the resistance of the insulating phase, it can still facilitate local filament formation at a lower critical voltage. Within this scenario, the memory effect is expected to become stronger at lower temperatures, as the switching voltage increases, thereby enhancing the electric field and the current surge following filament formation.
Since defect mobility is expected to be low up to 370K, thermal cycling up to this temperature is not expected to erase a defect-based memory effect, contrary to the metallic domains redistribution scenario. However, annealing at higher temperatures is expected to alter the defect distributions and therefore significantly modulate the memory effect. \vspace{6pt} 

In the third hypothesis, the "structural memory" scenario, the effect originates from a structural difference between the insulating state before and after filament formation. The formation of metallic domains within an insulating matrix is known to modify the local transition temperature at the edges of metallic domains which grow through the insulating matrix and subsequently contract (ramp reversal memory effect) \cite{vardi_rampreversal_2017}. Additionally, when the high-symmetry metallic structure transforms back into the low-symmetry monoclinic insulating phase, the films will adopt a certain microscopic twinning configuration. This configuration may be different from the equilibrium domain configuration occurring during the quasistatic cooling in the absence of current. This may result in higher-strain state in the region where the filament has formed, and may lower the energy barrier for subsequent filament formation, thereby reducing the switching voltage in later cycles. This type of memory effect is also expected to be erased by thermal cycling.
\vspace{6pt}

To distinguish between these hypotheses, we first investigated the temperature dependence of the memory effect, as different hypotheses predict a different response to temperature variation. We selected a range of temperatures along the R(T) cooling curve, covering points both inside and outside the hysteresis region. At each temperature, several devices were subjected to several consecutive IV measurement sweeps. From each sweep, the switching voltage was extracted, and the relative difference in switching voltage between the first and second sweeps was calculated. The cooling branch was chosen for the temperature dependence experiment because it contains a higher fraction of metallic domains than the equilibrium state. In this regime, any relaxation toward equilibrium would be expected to increase the number of insulating domains and thereby raise the switching voltage. Thus, a memory effect manifested as a \textit{decrease} in switching voltage cannot be attributed to equilibration but must instead reflect a new metastable state.\vspace{6pt}

\begin{figure}[!htbp]
     \centering
      \includegraphics[width=\linewidth]{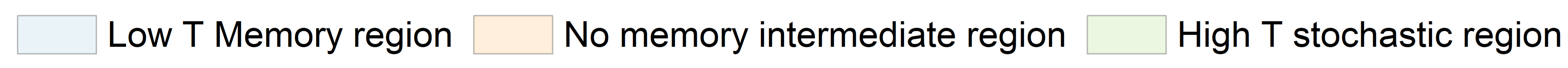}
     \begin{subfigure}[h]{0.48\textwidth}
         \centering
         \includegraphics[width=\textwidth]{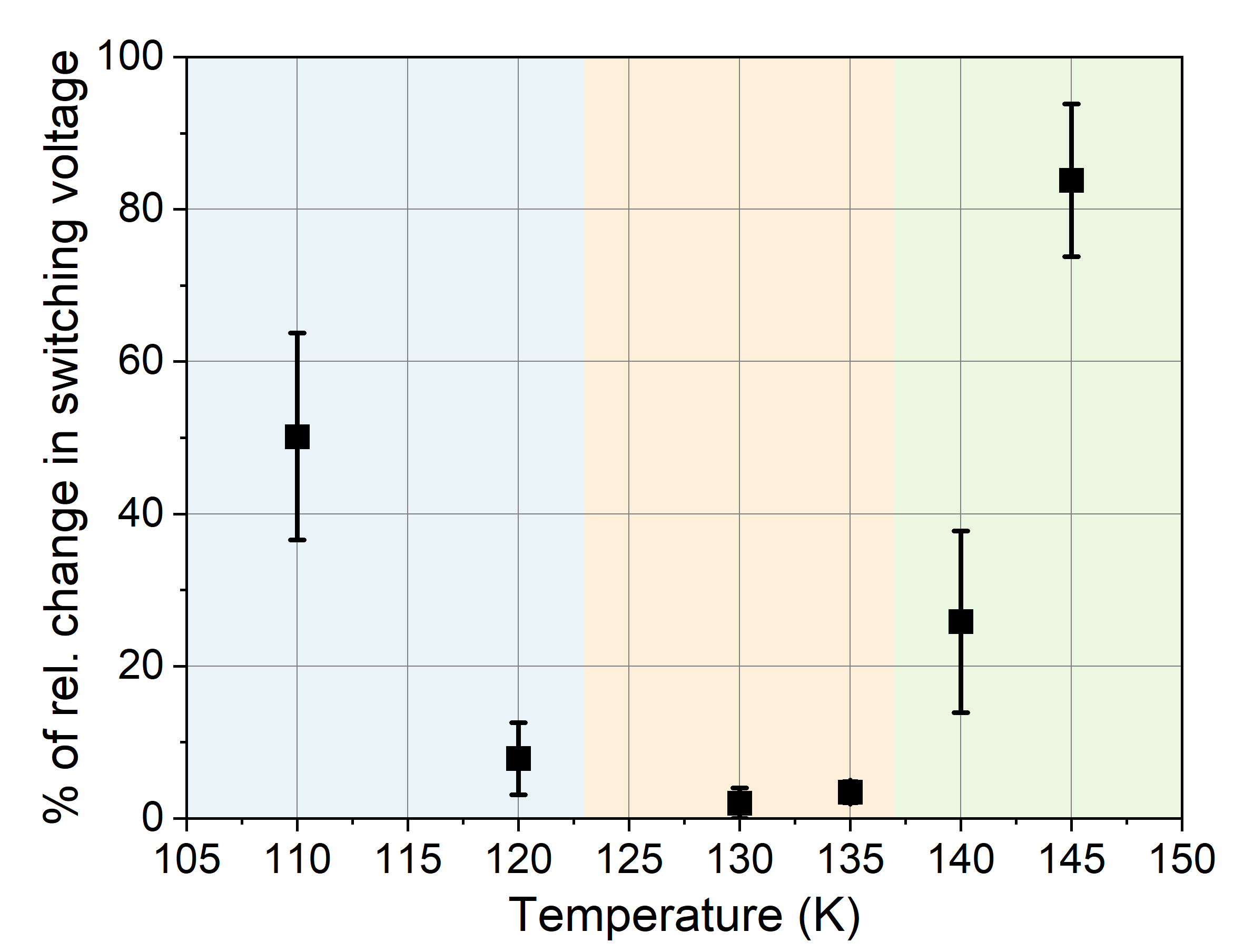}
         \caption{}
         \label{fig: dV V vs T V2O3}
     \end{subfigure}
     \hfill
     \begin{subfigure}[h]{0.48\textwidth}
         \centering
         \includegraphics[width=\textwidth]{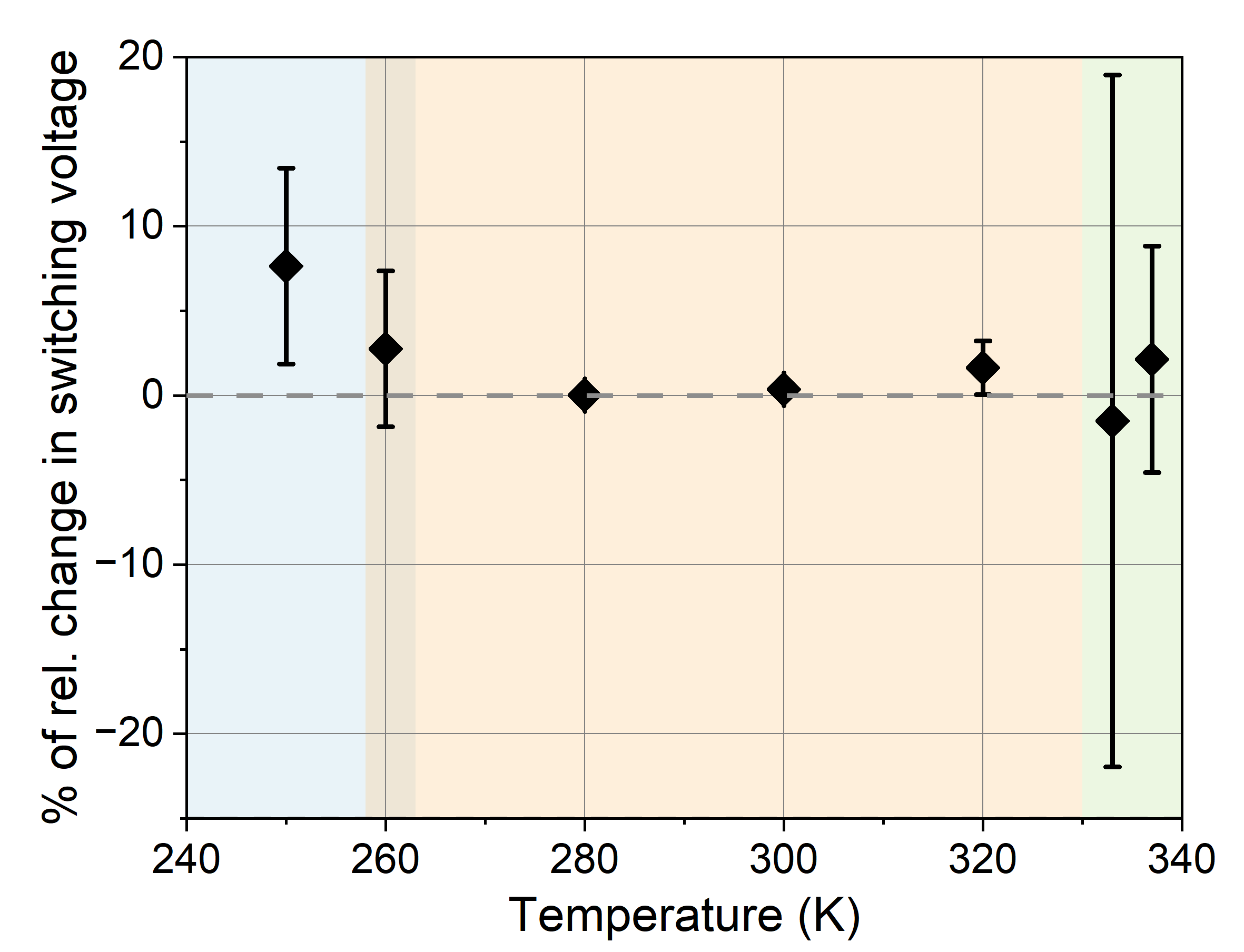}
         \caption{}
         \label{fig: dV V vs T VO2}
     \end{subfigure}
     \hfill
     \begin{subfigure}[h]{0.48\textwidth}
         \centering
         \includegraphics[width=\textwidth]{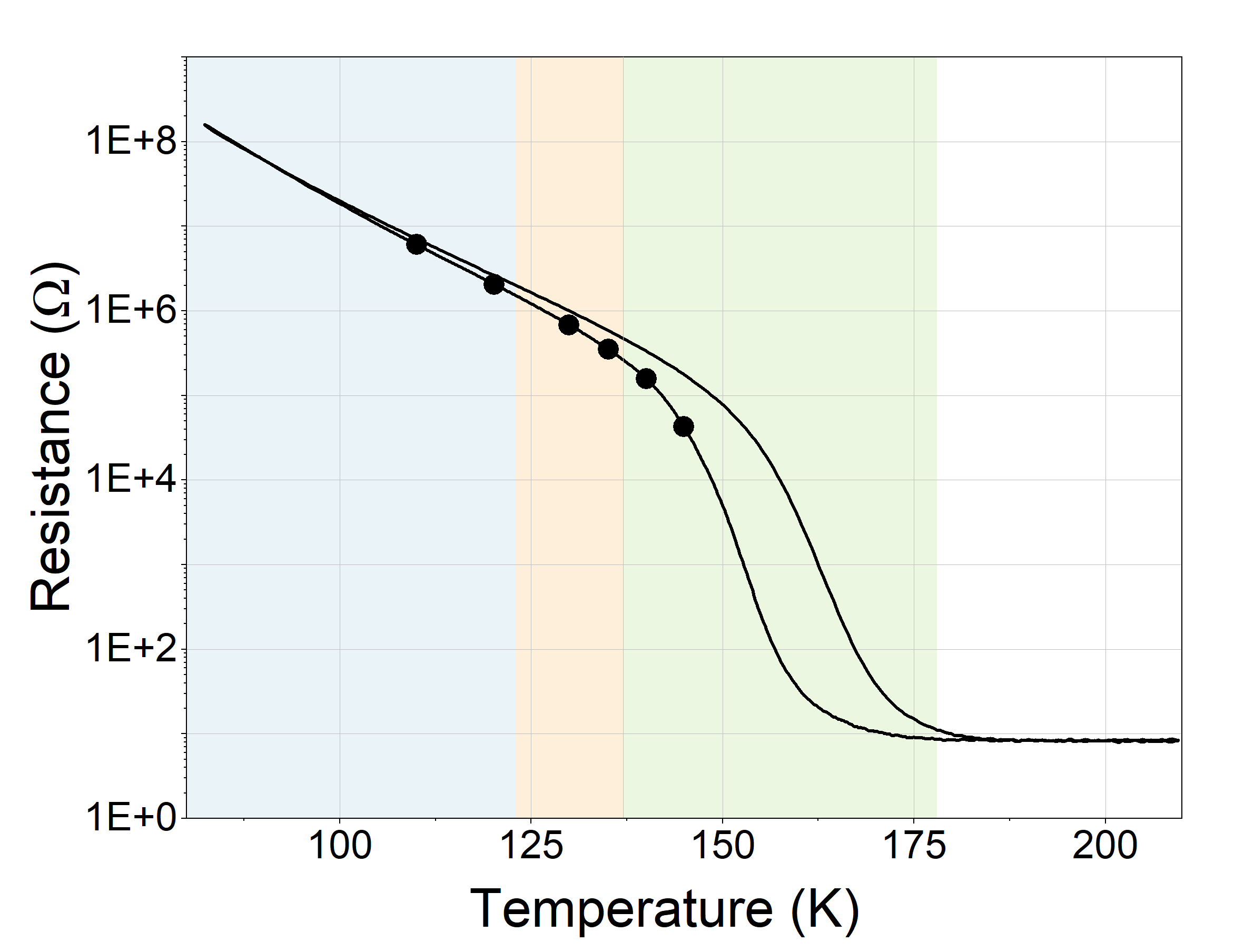}
         \caption{}
         \label{fig: dV V vs T V2O3 RT}
     \end{subfigure}
     \hfill
     \begin{subfigure}[h]{0.48\textwidth}
         \centering
         \includegraphics[width=\textwidth]{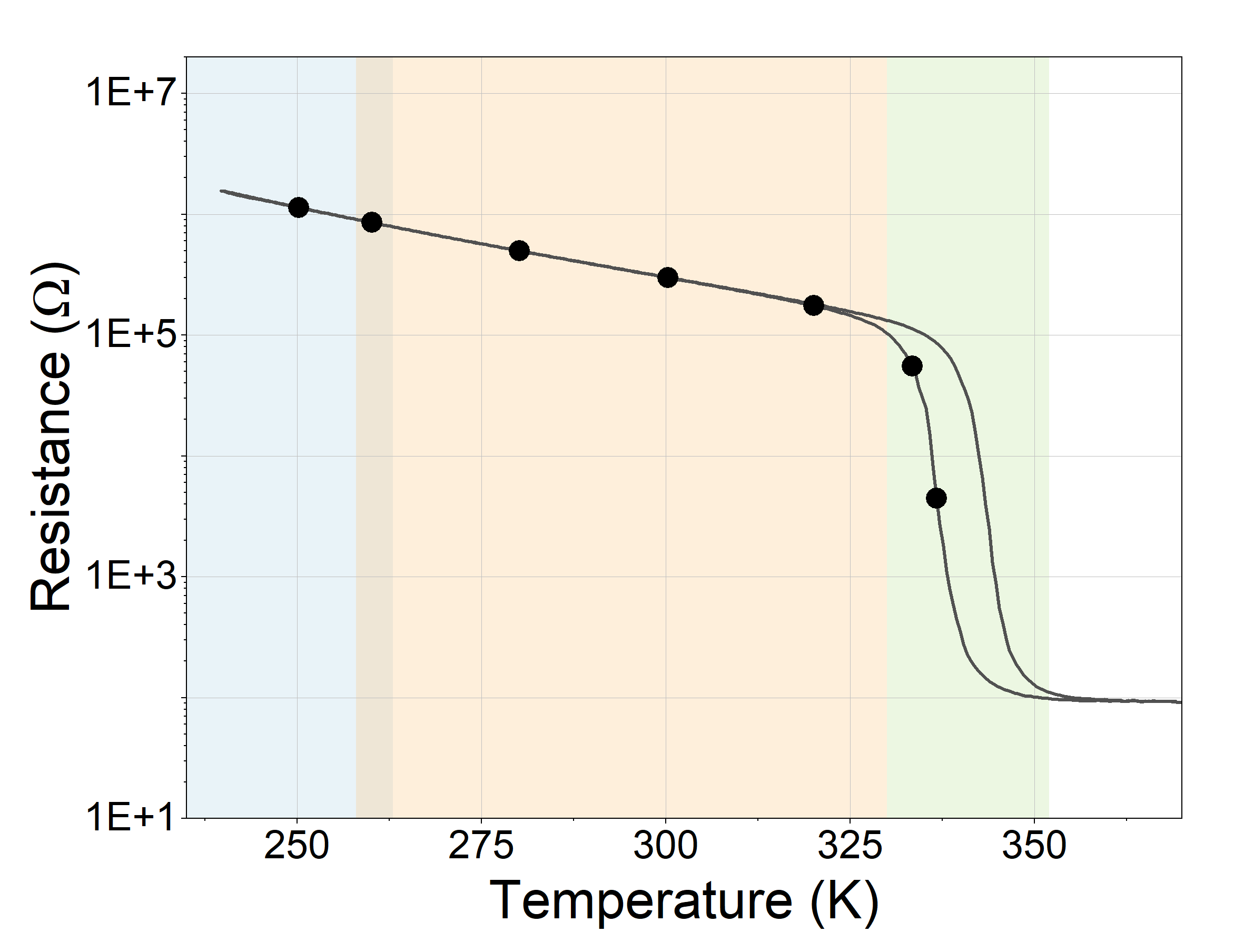}
         \caption{}
         \label{fig: dV V vs T VO2 RT}
     \end{subfigure}
     \hfill
    \caption{Relative change between the switching voltage of the 1st sweep and subsequent sweeps, defined as \(\frac{V_{sw,1}-V_{sw,2}}{V_{sw,2}}\), calculated for different temperatures for \textbf{(a)} V\textsubscript{2}O\textsubscript{3} and \textbf{(b)} VO\textsubscript{2}. \textbf{(c-d)} The RT curves of the respective films with the location on the cooling curve of the IV measurements temperatures highlighted. Shaded regions denote the different memory regimes, as indicated in the legend. Overlapping regions mark temperature ranges in which different devices exhibit different memory regimes}
    \label{fig:dv V vs T}
\end{figure}

The results of the experiment, which are shown in Figure \ref{fig:dv V vs T}, reveal three distinct temperature regimes, each characterized by a different memory behavior. We discuss these regimes sequentially, moving from high to low temperature.\vspace{6pt}

In the high temperature regime, which coincides with the phase-coexistence region and is indicated by a light green shaded region in Figure \ref{fig:dv V vs T}, switching is inherently stochastic and temperature-sensitive, because small changes in the domain configuration or measurement temperature can strongly affect the percolation of metallic domains and, consequently, the measured switching voltage \cite{cheng_inherent_2021}. Nevertheless, in V\textsubscript{2}O\textsubscript{3}, we observe a memory effect which becomes stronger with increasing temperature. Such a memory effect is consistent with the redistribution of metallic domains after the first switching event. In contrast, no such memory effect is observed in VO\textsubscript{2}.\vspace{6pt} 

At lower temperatures, just below the onset of the hysteresis curve and indicated by the light orange shaded region in Figure \ref{fig:dv V vs T}, there is an intermediate region, where no memory effect is observed in both materials. The existence of such a region allows us to discount the purely structural explanation for the phenomenon (i.e. the "structural memory" hypothesis). While the magnitude of residual strain following filament annihilation may be temperature-dependent, such structural memory would be expected to remain significant across the entire experimental range \cite{vardi_rampreversal_2017}. The existence of a regime where the disparity between the initial and subsequent IV sweeps vanishes suggests that a strain-based model is insufficient to account for the observed behavior. \vspace{6pt}

At even lower temperatures, indicated by the light blue shaded region in Figure \ref{fig:dv V vs T}, appears a region where a memory effect is observed for both materials, albeit with a different magnitude. In this region, the memory effect becomes stronger at lower temperatures, which is consistent with the defect-related mechanism.\vspace{6pt}

\begin{figure}[!htbp]
     \centering
     \includegraphics[width=0.7\linewidth]{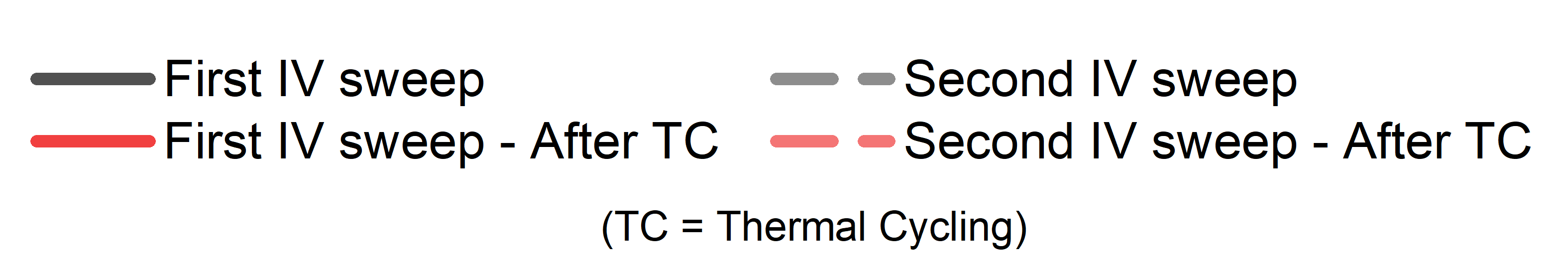}
     \begin{subfigure}[h]{0.48\textwidth}
         \centering
         \includegraphics[width=\textwidth]{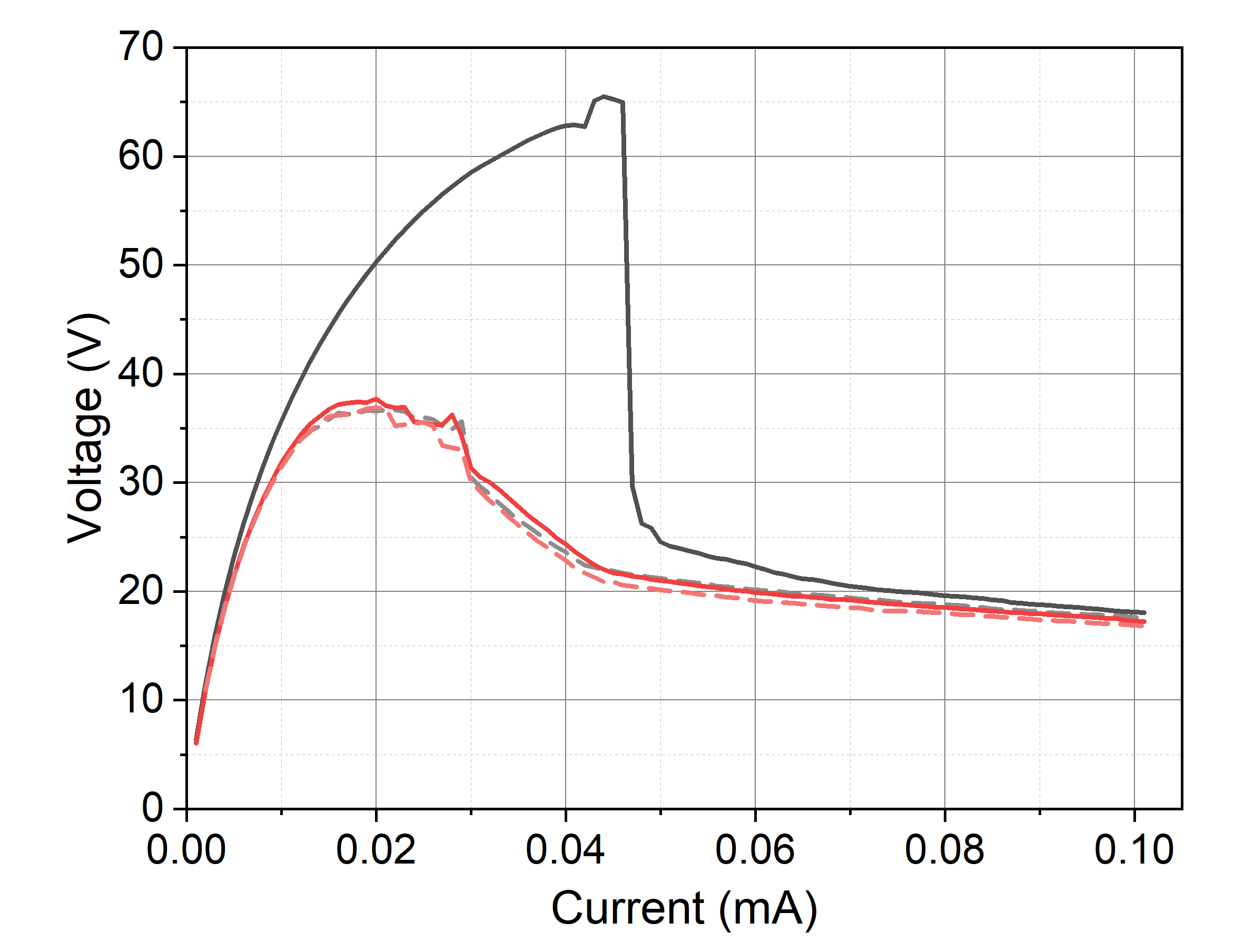}
         \caption{}
         \label{fig: Thermal cycling low T}
     \end{subfigure}
     \hfill
     \begin{subfigure}[h]{0.48\textwidth}
         \centering
         \includegraphics[width=\textwidth]{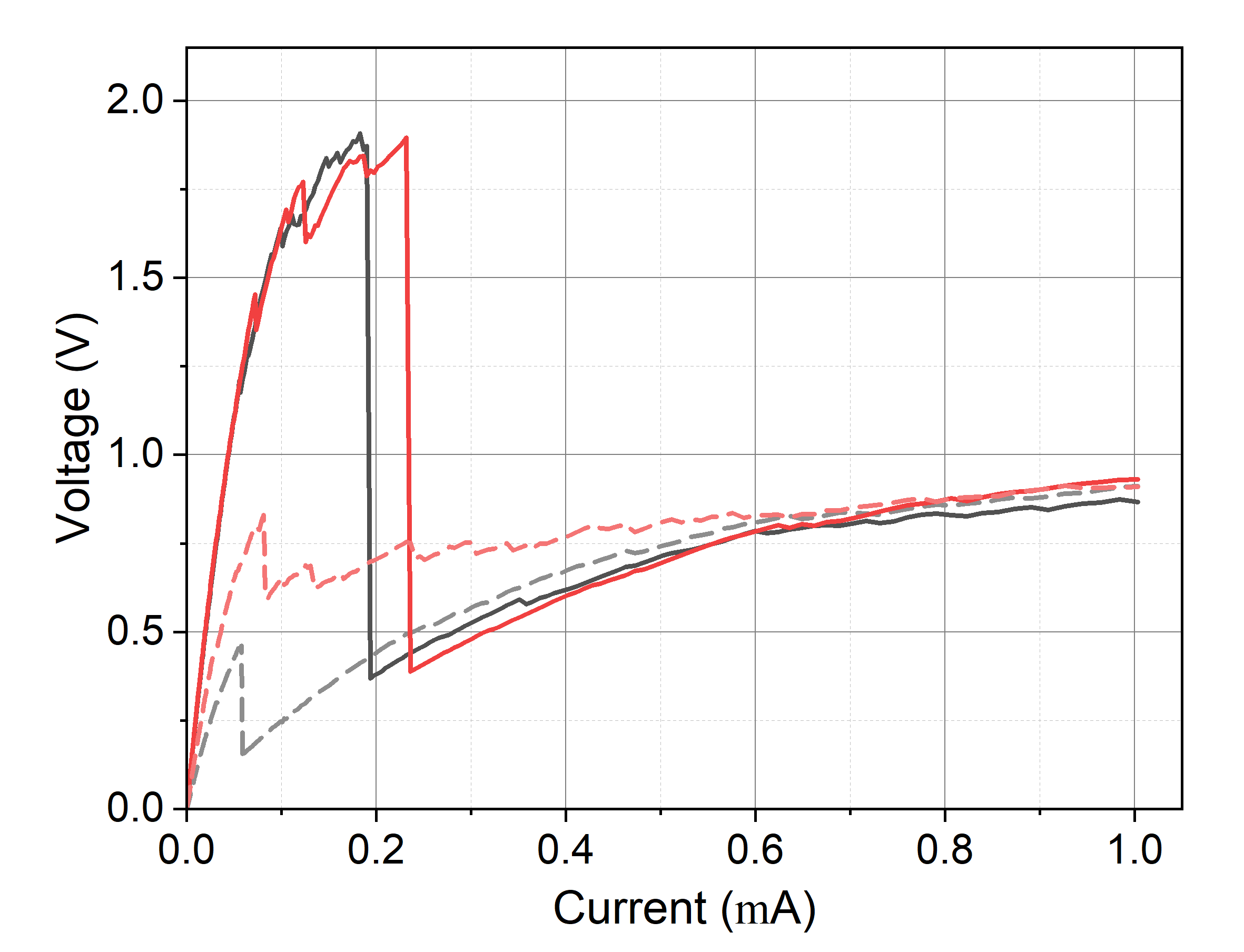}
         \caption{}
         \label{fig: Thermal cycling high T}
     \end{subfigure}
     \hfill
    \caption{Effect of thermal cycling on the observed memory effect at \textbf{(a)} low temperatures (T=110K) in the fully insulating state and at \textbf{(b)} high temperature (T=155K) in the phase coexistence regime in V\textsubscript{2}O\textsubscript{3} devices.}
    \label{fig: Thermal cycling effects}
\end{figure}

This behavior suggests the presence of two distinct memory effects with opposite temperature dependencies, arising from different physical mechanisms. To further support this interpretation, we investigated the influence of thermal cycling on the memory effects observed in the different temperature regimes of V\textsubscript{2}O\textsubscript{3}, as shown in Figure \ref{fig: Thermal cycling effects}.\vspace{6pt}

The response to thermal cycling provides a direct way to distinguish between the possible origins of the memory effect in V\textsubscript{2}O\textsubscript{3}. Within the phase-coexistence regime, thermal cycling erases the memory effect, as the switching voltage returns to its original value. By contrast, at low temperatures, thermal cycling has no measurable influence on the switching voltage, and the reduced switching voltage persists. This persistence is incompatible with the metallic domains redistribution and the structural-modification hypotheses, in which the memory effect is expected to be affected by thermal cycling. Instead, the low-temperature behavior is consistent only with a defect-based mechanism.\vspace{6pt}

\begin{figure}[!htbp]
\centering
\includegraphics[width=0.9\textwidth]{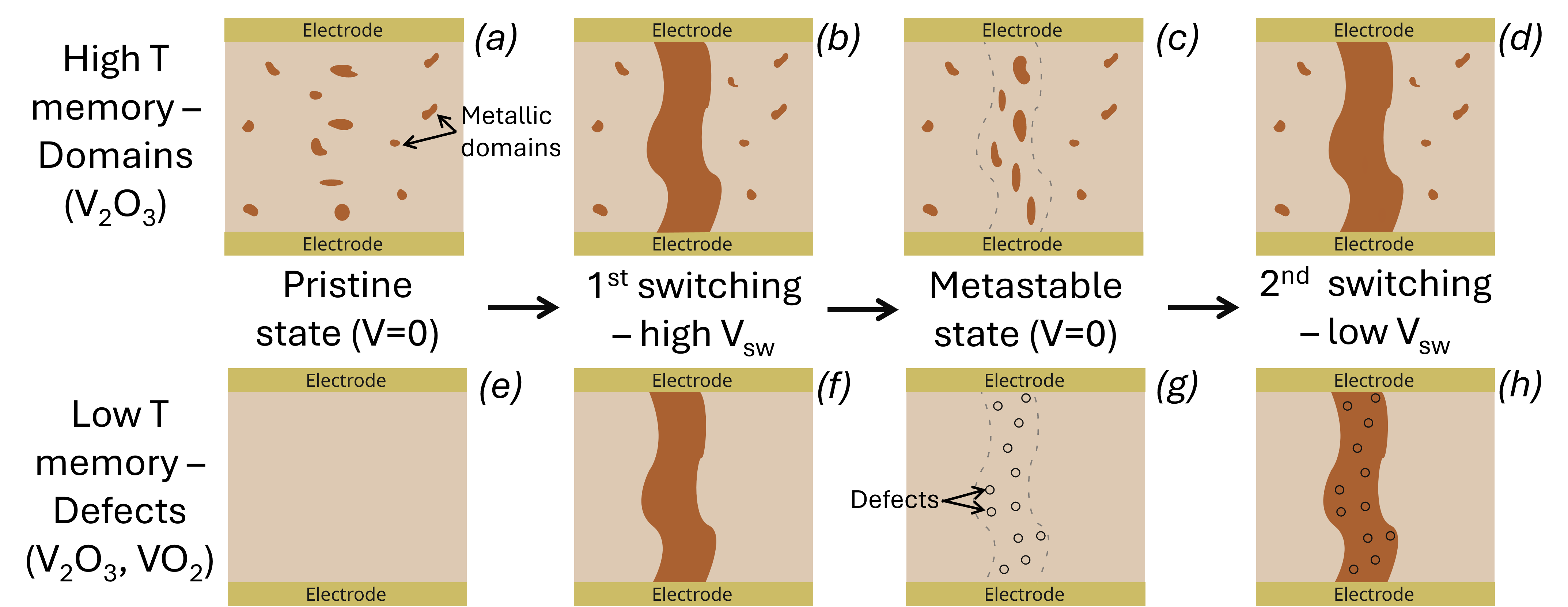}
\caption{Schematic of the mechanism of two memory effects observed in VO\textsubscript{x} devices: \textbf{(a-d)} The domain redistribution memory - the pristine state contains a certain fraction of metallic domains, following the 1st IV, the domains rearrange in the filament path, lowering the voltage required for the re-creation of the filament.  \textbf{(e-h)} The defects based memory - following the first switching event, defects are concentrated in the filament path, lowering the voltage required for the re-creation of the filament.} 
\label{fig: mechanisms schematic}
\end{figure}

Taken together, these results reveal two distinct memory mechanisms in vanadium oxide VRS devices. In both V\textsubscript{2}O\textsubscript{3} and VO\textsubscript{2}, a defect-related memory effect appears at low temperatures within the fully insulating state. Additionally, in V\textsubscript{2}O\textsubscript{3} only, there is also a memory effect inside the phase-coexistence regime associated with the redistribution of metallic and insulating domains. These mechanisms and their corresponding temperature regimes are summarized schematically in Figure \ref{fig: mechanisms schematic}.
\vspace{6pt}

Having established that the low-temperature and phase-coexistence memory effects arise from distinct mechanisms, the remainder of the main text focuses on the defect-related memory effect. This effect is particularly important because it enables controllable modification of the switching voltage and can therefore be exploited for post-fabrication tuning of device properties. Additional aspects of the phase-coexistence memory effect, mainly its robustness against the inherent stochasticity of switching in the phase-coexistence regime, are discussed in detail in supplementary section S5. \vspace{6pt}

\begin{figure}[!htbp]
     \centering
     \begin{subfigure}[h]{0.48\textwidth}
         \centering
         \includegraphics[width=\textwidth]{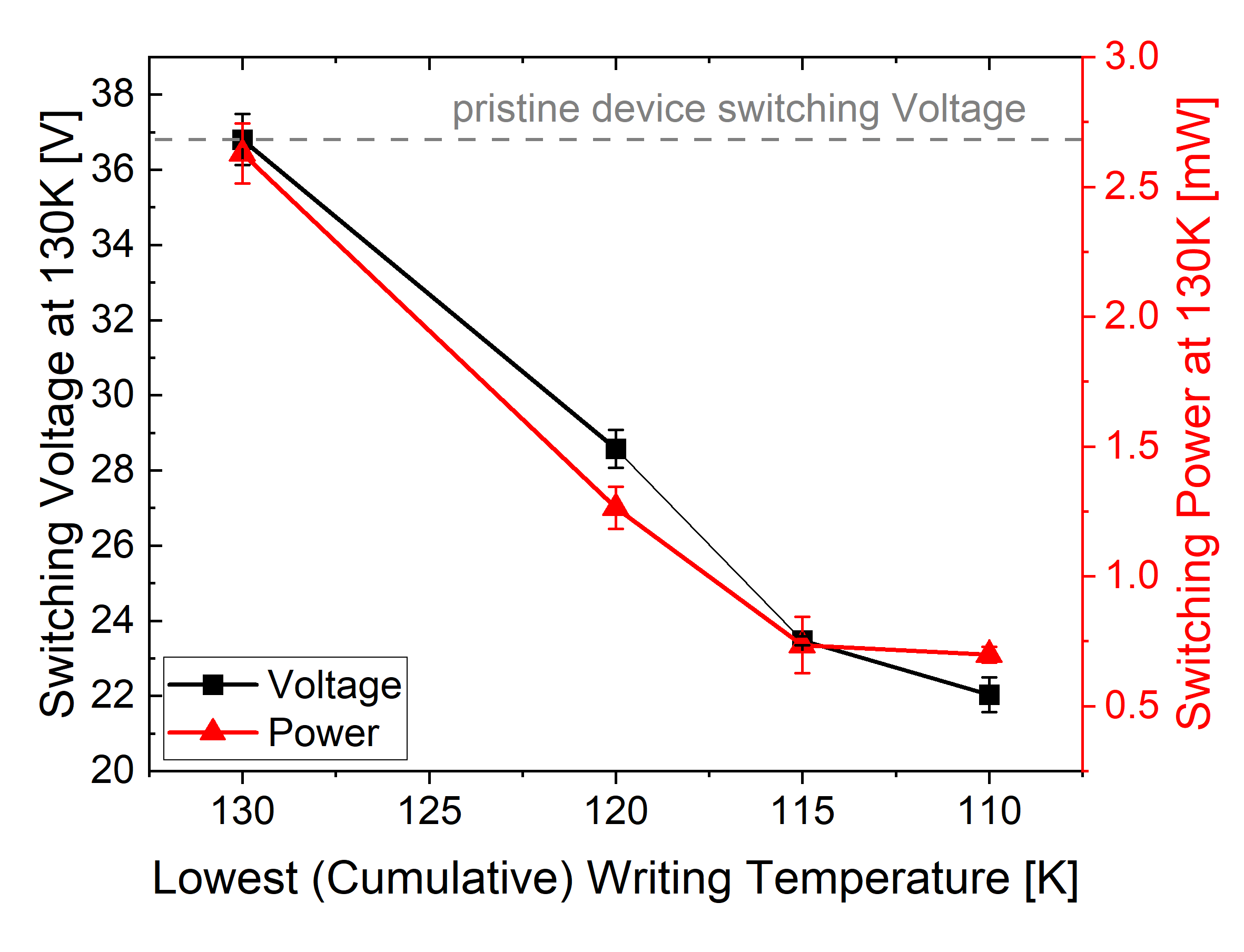}
         \caption{}
         \label{fig: IV writing V2O3}
     \end{subfigure}
     \hfill
     \begin{subfigure}[h]{0.48\textwidth}
         \centering
         \includegraphics[width=\textwidth]{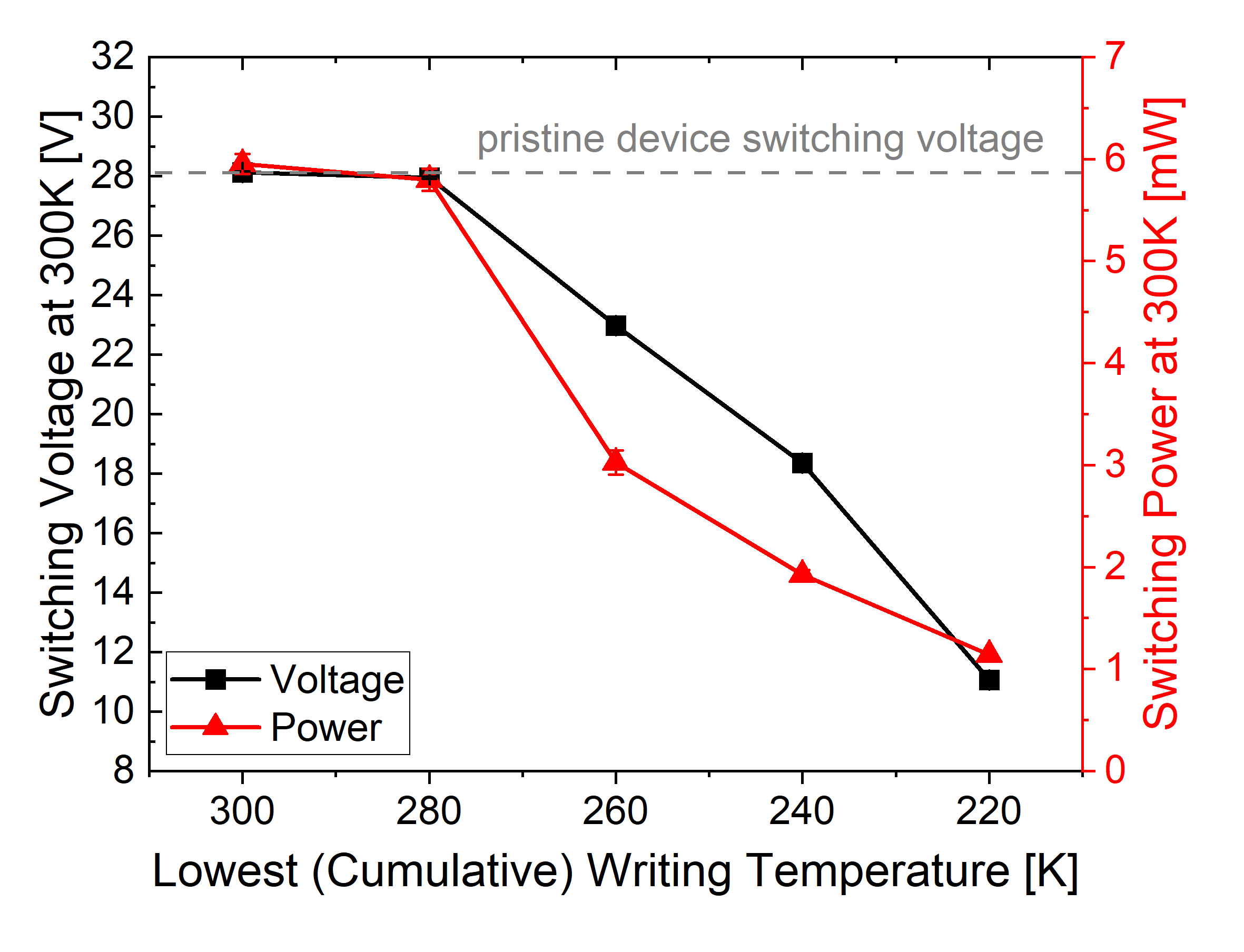}
         \caption{}
         \label{fig: IV writing VO2}
     \end{subfigure}
     \hfill
    \caption{The switching voltage and power at the operating temperature after each step of measuring IV at a lower temperature ('writing') in \textbf{(a)} V\textsubscript{2}O\textsubscript{3} and \textbf{(b)} VO\textsubscript{2}. The writing procedure is cumulative: each writing step is performed after all preceding writing steps at higher temperatures. The error bars are the standard deviation of the switching voltage (power) between N$\geq$4 consecutive sweeps.}
    \label{fig: IV writing Exp}
\end{figure}

The defect-related low-temperature memory effect is not merely a consequence of the switching process, but can also serve as a practical mechanism for controlling device parameters. The controlled introduction of defects can be leveraged to tune the switching voltage at a pre-chosen operating temperature, even one in which defect-related memory effects are otherwise negligible. The tuning procedure is as follows: an initial IV sweep is recorded at an operating temperature to determine the initial switching parameters, after which the device is cooled to a lower temperature for a subsequent IV measurement (the 'writing step'). The device is then returned to the operating temperature while remaining on the cooling curve, achieved by heating it into the fully metallic state and then re-cooling, before a second IV measurement is performed. Figure \ref{fig: IV writing Exp} shows the switching voltage and power that were measured at the operating temperature after each writing step. The operating temperature lies in the intermediate region, where no memory effect was observed, yet the switching voltage and power were systematically reduced once the writing was performed below a threshold temperature. This threshold is closely linked to the defect-related memory, as the switching parameters began to decrease only when the writing temperature entered the regime in which defect-related memory effects were measurable (as shown in Figure \ref{fig: dV V vs T VO2}). \vspace{6pt}

\begin{figure}[!htbp]
     \centering
     \begin{subfigure}[h]{0.48\textwidth}
         \centering
         \includegraphics[width=\textwidth]{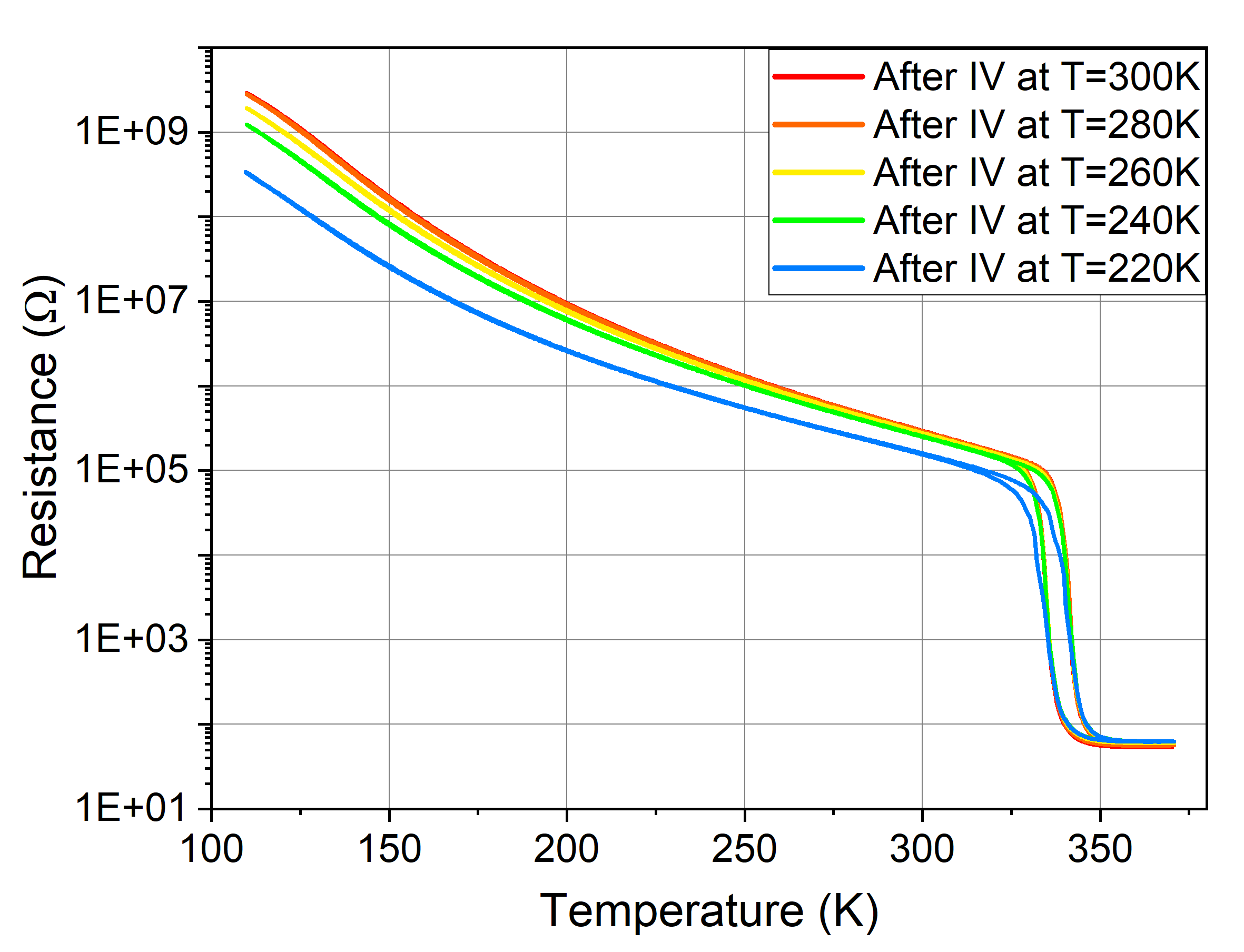}
         \caption{}
         \label{fig: RT gradual writing}
     \end{subfigure}
     \hfill
     \begin{subfigure}[h]{0.48\textwidth}
         \centering
         \includegraphics[width=\textwidth]{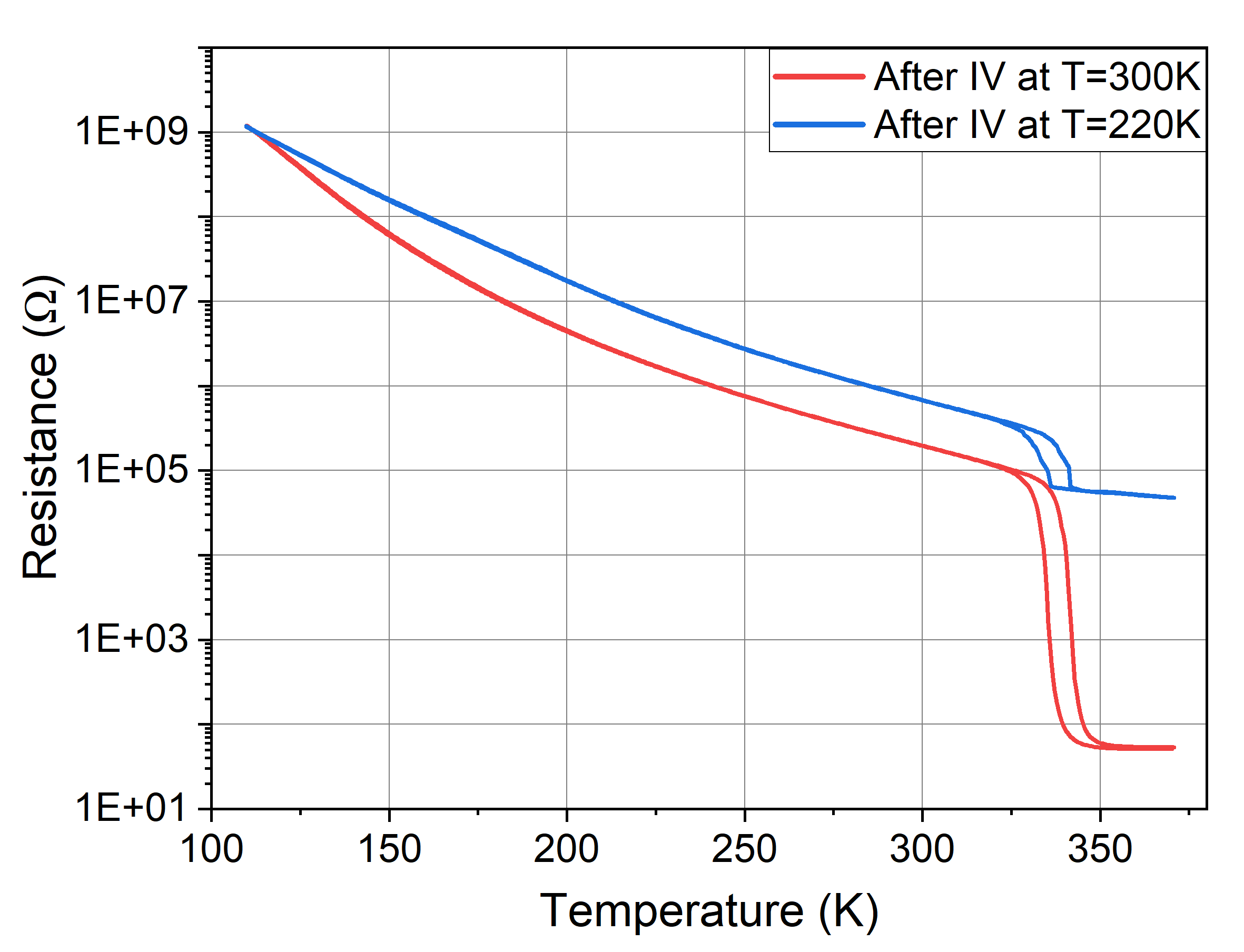}
         \caption{}
         \label{fig: RT sudden writing}
     \end{subfigure}
     \hfill
     \vspace{6pt}
     \begin{subfigure}[h]{0.48\textwidth}
         \centering
         \includegraphics[width=\textwidth]{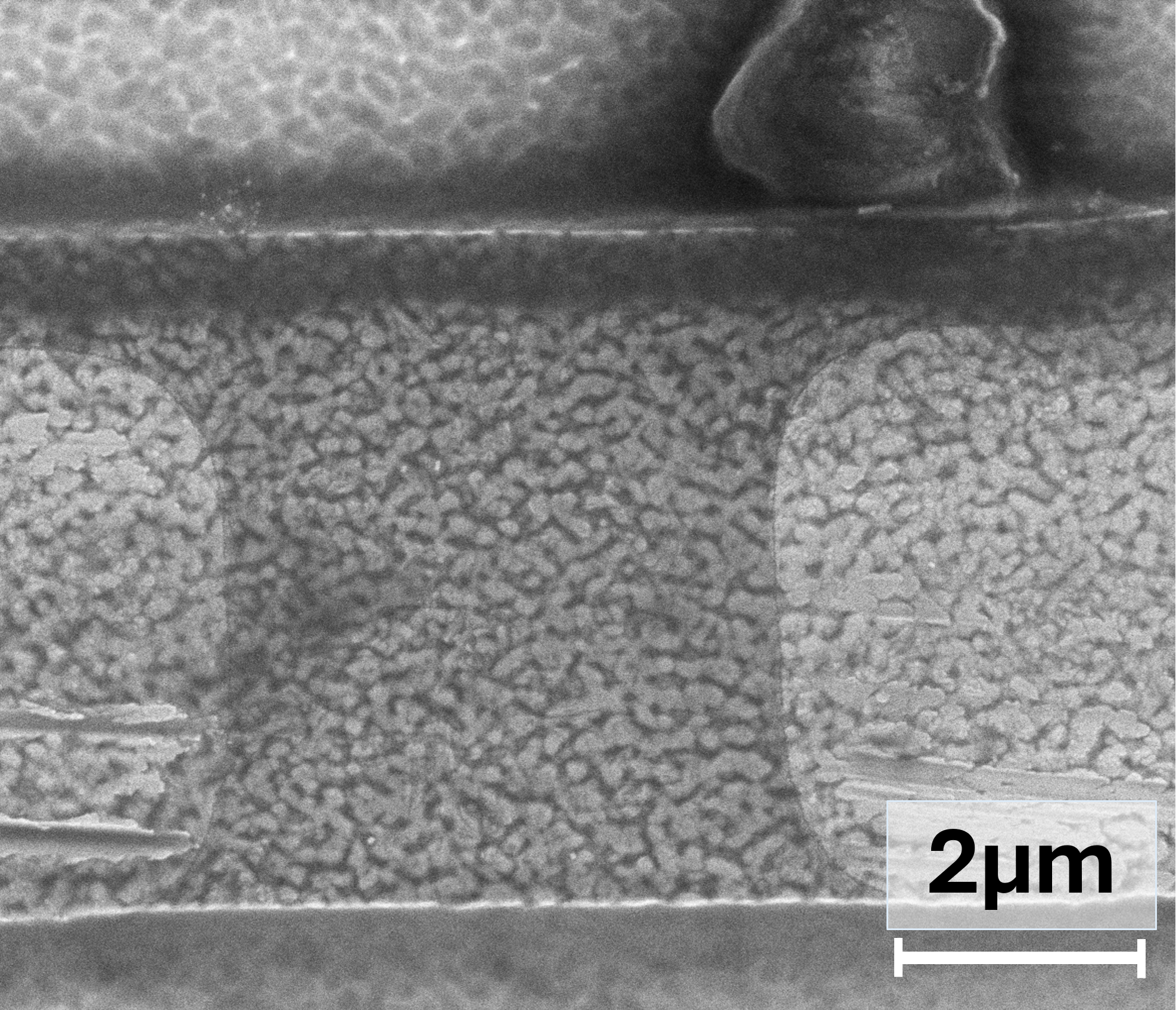}
         \caption{}
         \label{fig: SEM gradual writing}
     \end{subfigure}
     \hfill
     \begin{subfigure}[h]{0.48\textwidth}
         \centering
         \includegraphics[width=\textwidth]{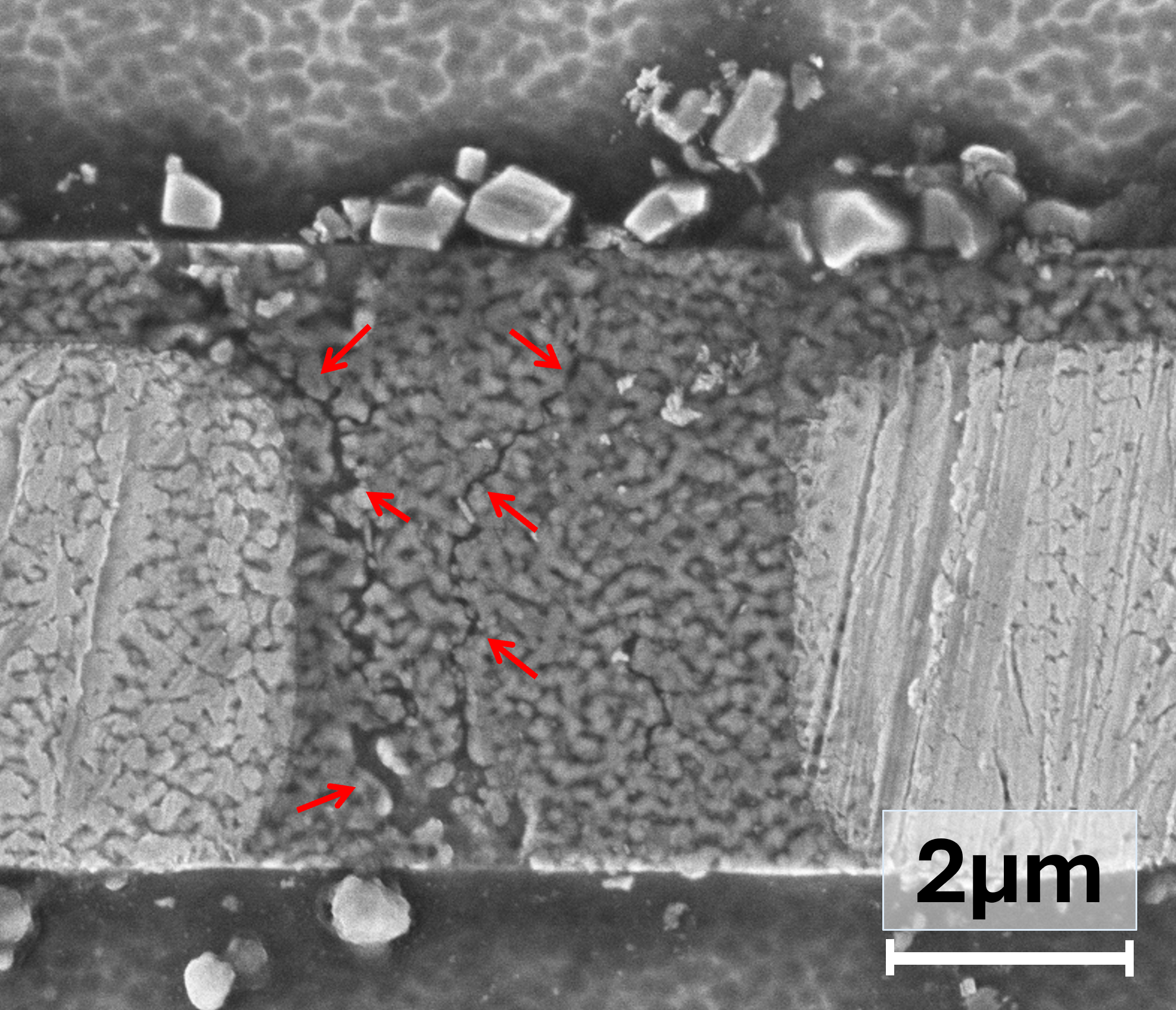}
         \caption{}
         \label{fig: SEM sudden writing}
     \end{subfigure}
     \hfill
    \caption{\textbf{(a-b)} Comparison of the R(T) curves of a VO\textsubscript{2} device before and after \textbf{(a)} cumulative writing procedure and \textbf{(b)} sudden writing procedure. The red curve in \textbf{(a)} overlap with the orange one, as no change in the RT curve was measured after this writing step. \textbf{(c-d)} SEM micrographs of the respective VO\textsubscript{2} devices. The red arrows in \textbf{(d)} highlight the path of the two longest cracks that were formed in the device following the sudden writing procedure.}
    \label{fig: RT and SEM after writing}
\end{figure}

The gradual nature of the writing protocol is crucial for controlling the switching voltage. As shown in Figure \ref{fig: IV writing VO2}, when the writing steps were performed gradually from 300K to 220K in 20K intervals, the switching voltage was reduced by a factor of $\sim$2.5 and the switching power by a factor of $\sim$6, accompanied only by a mild decrease in the insulating-phase resistance measured from the RT curve (Figure \ref{fig: RT gradual writing}). By contrast, a sudden writing step at 220K produced a markedly different RT curve, where the resistance of both phases \textit{increased}, including an orders-of-magnitude increase in the metallic-phase resistance (Figure \ref{fig: RT sudden writing}). \vspace{6pt}

This difference suggests that the current surge affects the device in a fundamentally different manner when its magnitude is larger, as in the sudden writing protocol. Microscopic characterization by Scanning Electron Microscopy (SEM) reveals that the larger current surge produces structural damage in the device. As shown in Figure \ref{fig: SEM sudden writing}, the sudden writing protocol leads to the formation of cracks oriented parallel to the electrodes. These cracks partially disconnect the conducting path and therefore substantially increase the device resistance. By contrast, no such cracks are observed after the gradual writing protocol (Figure \ref{fig: SEM gradual writing}), and the device morphology remains similar to that of a pristine device (Supplementary figure S6).

Similar morphological changes could not be observed in V\textsubscript{2}O\textsubscript{3}, probably due to the limit on the lowest temperature accessible by our measurement setup, which could not exceed 200V.  \vspace{6pt} 

\begin{figure}[!htbp]
\centering
\includegraphics[width=0.8\textwidth]{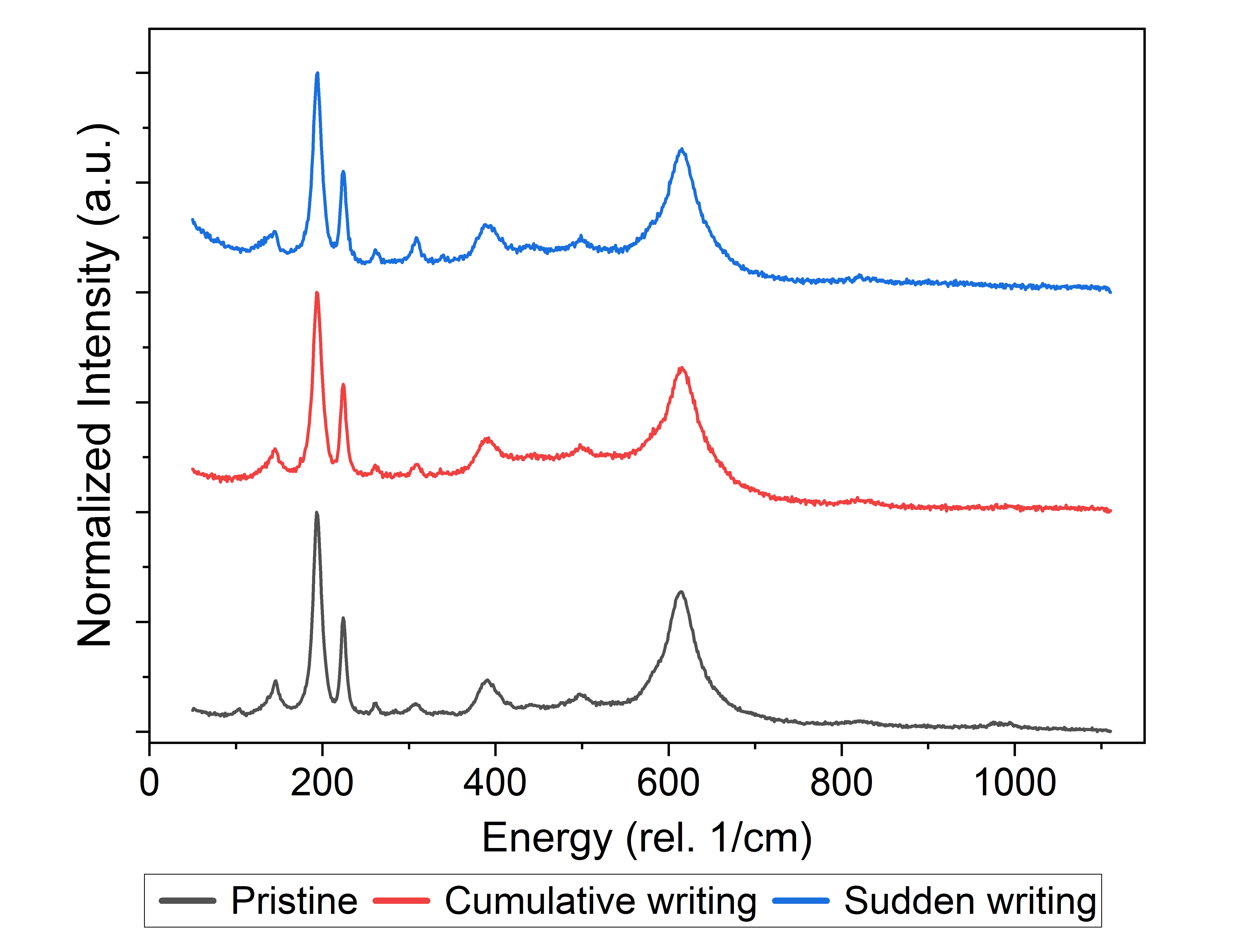}
\caption{Comparison of the Raman spectra of a pristine device and devices after cumulative writing and sudden writing procedures.}
\label{fig: Raman}
\end{figure}

The drastic change in the RT curve after the sudden writing protocol is therefore most likely associated with structural damage rather than the formation of a new, highly resistive oxide phase. This interpretation is further supported by the absence of additional hysteresis features in the RT curve, which would be expected if distinct VO\textsubscript{x} phases were formed following the IV sweep, as previously observed \cite{del_valle_electrically_2017}. Moreover, Raman measurements, shown in Figure \ref{fig: Raman}, reveal no significant spectral differences between devices subjected to gradual and sudden writing protocols. In both cases, all observed peaks are consistent with the monoclinic M1 phase of VO\textsubscript{2} \cite{shvets_review_2019}.

\section*{Discussion}

The observations presented here reveal a regime-dependent relation between volatile filament dynamics and non-volatile material modification in vanadium oxides.

Discussing first the phase-coexistence regime in V\textsubscript{2}O\textsubscript{3}, a similar memory effect has been reported previously \cite{guenon_electrical_2013,lange_imaging_2021}. However, in those cases, the initial IV sweep was performed on the heating branch of the hysteresis curve. On this branch, the current ramp-down effectively cools the device, moving it closer to the cooling curve, which corresponds to a higher fraction of metallic domains and therefore a lower resistance. As a result, the observed reduction in switching voltage is expected to occur. In contrast, the present results demonstrate a memory effect while the device remains on the cooling curve throughout all IV sweeps, this suggests a more subtle reconfiguration of the metallic and insulating domains rather than an increase in the metallic phase fraction. Previous direct imaging of domain dynamics in vanadium oxides has characterized the evolution of the IMT by various methods \cite{lange_imaging_2021,alzate_banguero_optical_2025,madan_quantitative_2015,kisiel_high-resolution_2025, tiwari_near_2026}. Applying similar imaging techniques before and after an IV sweep on the cooling curve would help clarify the exact spatial redistribution of domains following filament annihilation. Such measurements could also provide insight into possible differences in domain relaxation dynamics between VO\textsubscript{2} and V\textsubscript{2}O\textsubscript{3}, which may explain why a phase-coexistence memory effect is observed in V\textsubscript{2}O\textsubscript{3} but not in VO\textsubscript{2}. \vspace{6pt}

We now turn to the low-temperature defect-related memory effect, which should be understood within the broader context of surge-induced device modification. At the highest surge magnitudes, which occur in the absence of in-situ device protection, the device undergoes catastrophic modification, such as local melting or formation of other oxides. The findings of previous studies of electroforming in VO\textsubscript{2} devices \cite{conti_electroforming_2026, del_valle_electrically_2017,cheng_operando_2021}, as well as the device melting we observed in a device without protection (Supplementary Section S2), belong to this regime.  At lower, but still substantial, surge magnitudes, melting is avoided, but structural damage such as crack formation can occur \cite{giorgianni_overcoming_2019}. In both cases, the device properties are largely determined by a single irreversible damage event, and the subsequent switching behavior reflects the altered device structure.\vspace{6pt}

A qualitatively different behavior emerges in the milder defect-forming regime. Here, the current surge does not catastrophically damage the device, but instead introduces a controllable concentration of defects. These defects modify the switching voltage, which in turn changes the magnitude of the current surge in subsequent IV sweeps. This creates a feedback loop between volatile filament formation and non-volatile defect generation: each switching event produces a mild current surge that introduces defects, which then lower the switching voltage for subsequent switching events at lower temperatures. The reduced switching voltage at low temperature keeps the following surge mild, resulting in a gradual increase of defects concentration without inducing more severe device damage. This interplay between volatile switching and non-volatile material modification transforms the electroforming from a single aggressive event into a controlled sequence of gradual forming steps.\vspace{6pt} 

Although the microscopic nature of the defects in this regime is unclear, oxygen vacancies are the most plausible candidates. Oxygen vacancies are known to form in vanadium oxides subjected to elevated temperatures in oxygen-poor environments \cite{xu_effects_2016,lee_controlling_2021}, with recent studies in VO\textsubscript{2} showing that an oxygen deficient patch is formed after RS \cite{tiwari_near_2026}. Additional support for the involvement of oxygen-related defects comes from the qualitatively similar changes in the electrical properties we observed in devices subjected to mild annealing treatments (Supplementary section S7), and reported in literature \cite{zhang_evolution_2017_PRapplied,zhang_evolution_2017}. Thus, rather than producing a distinct secondary phase, gradual electroforming may lead to the formation of defective VO\textsubscript{2-$\delta$} or V\textsubscript{2}O\textsubscript{3-$\delta$} phases. This interpretation, however, remains to be confirmed by nanoscale chemical and structural characterization. Such measurements are challenging because the relevant changes are expected to involve a localized filamentary region and potentially small variations in oxygen-vacancy concentration, requiring both high spatial resolution and high chemical sensitivity.\vspace{6pt}

It is worth noting that the ability to tune the switching voltage through defect-mediated memory is not necessarily limited to vanadium oxides, this approach could be equally relevant to other Mott insulators such as NbO\textsubscript{2}, where control over the switching voltage is a critical factor for device application.\vspace{6pt}

Moreover, the defect-related memory effect observed here shares significant similarity with the 'first-firing voltage' reported in Chalcogenide glasses exhibiting Ovonic switching behavior \cite{kim_firing_2020,kim_nanoscopic_2026}. In those systems, the effect also intensifies at lower temperatures \cite{kim_boosting_2025}. While the origin of the first-firing phenomenon is not yet fully elucidated, its phenomenological similarity to electroforming suggests that gradual electroforming protocols, such as the one introduced here for IMT-based VRS devices, may also be relevant for Ovonic Switching devices.\vspace{6pt}

\section*{Conclusions}

In summary, the present findings uncover regime-dependent memory effects in V\textsubscript{2}O\textsubscript{3} and VO\textsubscript{2} thin-film switching devices. These memory effects go beyond the expected behavior of a standard hysteretic system. It is found that nominally volatile resistive switching can be modified by non-volatile processes occurring during the switching event, thereby altering the subsequent volatile switching properties of the device. These processes give rise to two distinct memory mechanisms, in which the first switching event requires higher voltage than subsequent events. At temperatures within the phase-coexistence regime, the memory is associated with a spatial redistribution of metallic and insulating domains and occurs only in V\textsubscript{2}O\textsubscript{3}. However at low-temperatures, within the fully insulating phase, both oxides show a defect-mediated memory effect. This memory effect is understood as part of a broader hierarchy of current surge-induced device modifications. At the highest current-surge magnitudes, switching can lead to destructive changes such as local melting or crack formation. At lower surge magnitudes, the device remains structurally intact, but the surge can introduce defects that persistently modify the switching voltage. In this regime, the current surge may not constitute a destructive failure mechanism, but can be harnessed to tune switching parameters through control over defects.

From an application perspective, this defect-mediated memory effect provides a robust mechanism for post-fabrication device tuning. We demonstrate that a low-temperature writing process can be used to electrically modulate the switching voltage and power of VO\textsubscript{x}-based devices over a broad temperature range with minimal effect on the HRS/LRS ratio of the device. These findings establish controlled defect introduction as a promising pathway for tunable resistive-switching electronics.

\section*{Acknowledgments}

This work was funded by the European Union’s Horizon Europe research and innovation program (ERC, MOTTSWITCH, 101039986). Views and opinions expressed are however those of the authors only and do not necessarily reflect those of the European Union or the European Research Council Executive Agency. Neither the European Union nor the granting authority can be held responsible for them. This research was also supported by the Israel Science Foundation (grant no. 1031/21).

\bibliography{references}

\end{document}

\maketitle

\tableofcontents
\clearpage

\section{X-Ray Diffraction Pattern of the as-grown Films}
\begin{figure}[!htbp]
     \centering
     \begin{subfigure}[h]{0.48\textwidth}
         \centering
         \includegraphics[width=\linewidth]{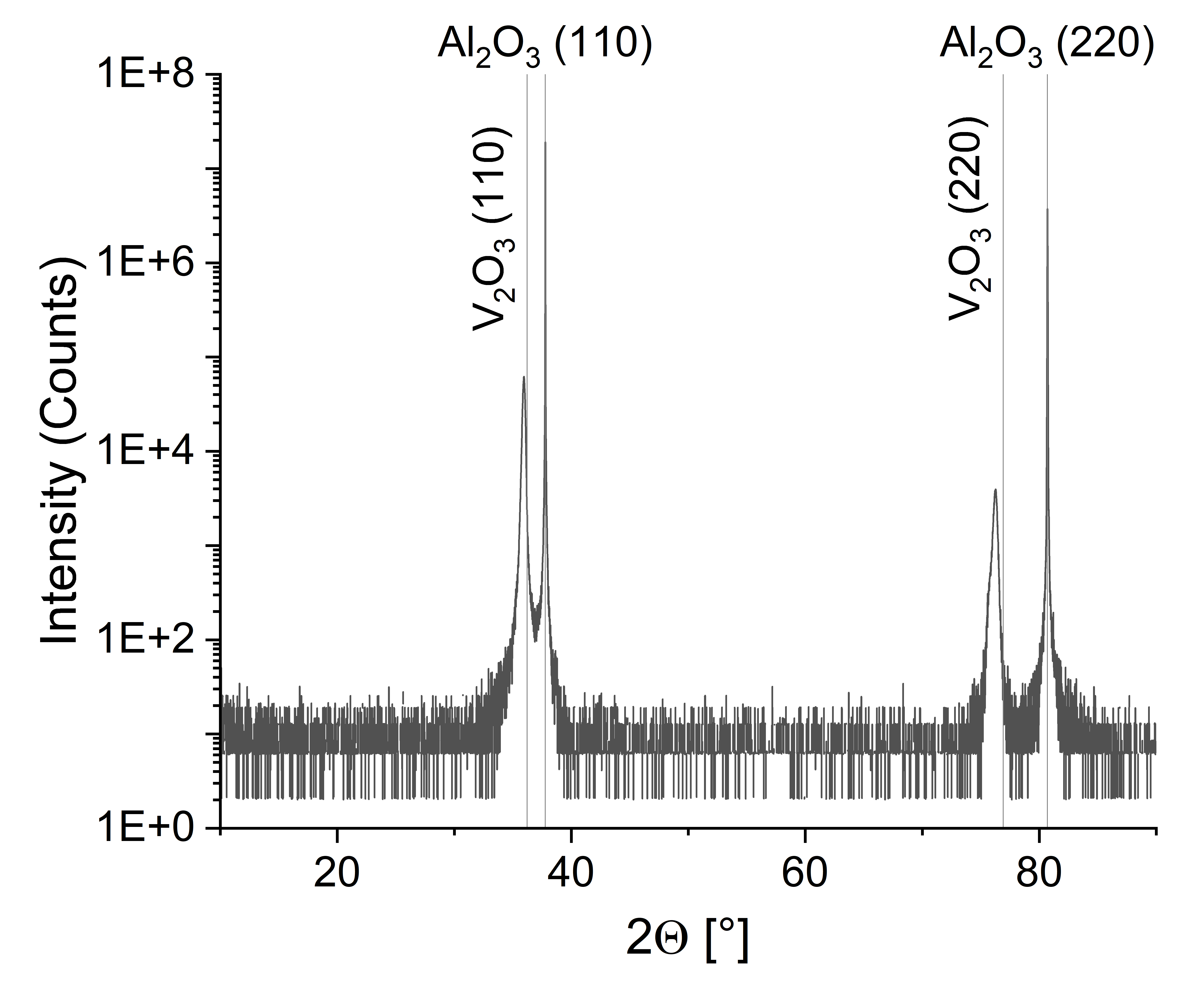}
         \caption{}
         \label{sup fig: XRD V2O3}
     \end{subfigure}
     \hfill
     \begin{subfigure}[h]{0.48\linewidth}
         \centering
         \includegraphics[width=\textwidth]{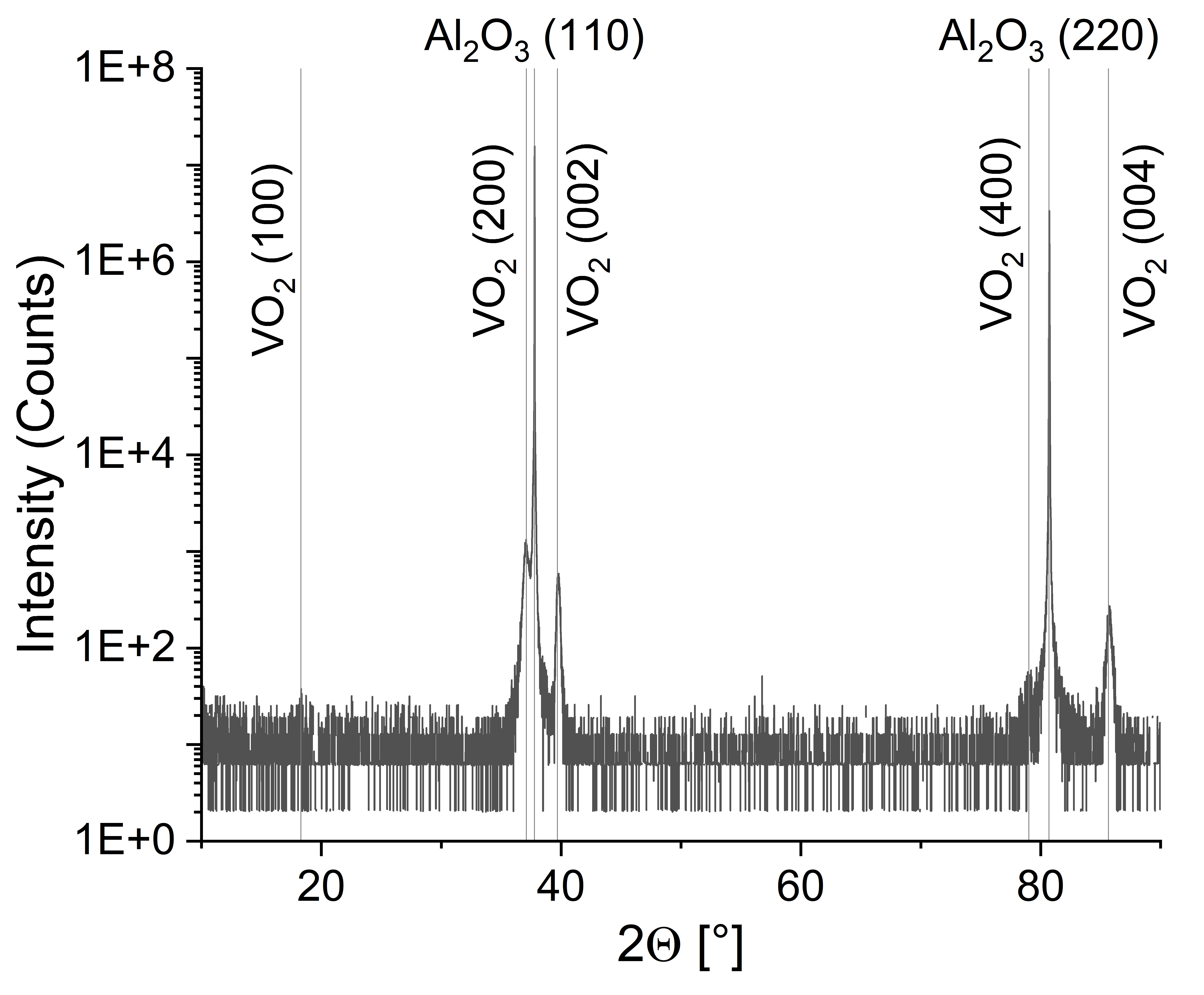}
         \caption{}
         \label{sup fig: XRD vo2}
     \end{subfigure}
     \hfill
    \caption{XRD of the as-grown \textbf{(a)} V\textsubscript{2}O\textsubscript{3} and \textbf{(b)} VO\textsubscript{2} thin-films.} 
    \label{sup fig: XRD VOx}
\end{figure}

X-ray diffraction patterns for pristine V\textsubscript{2}O\textsubscript{3} and VO\textsubscript{2} films grown on A-cut sapphire are shown in Figure \ref{sup fig: XRD VOx}. Both films are single-phase and highly oriented.

\section{Damage Induced by the Current Surge to a Device without Protection}

\begin{figure}[!htbp]
     \centering
     \begin{subfigure}[h]{0.48\textwidth}
         \centering
         \includegraphics[width=\linewidth]{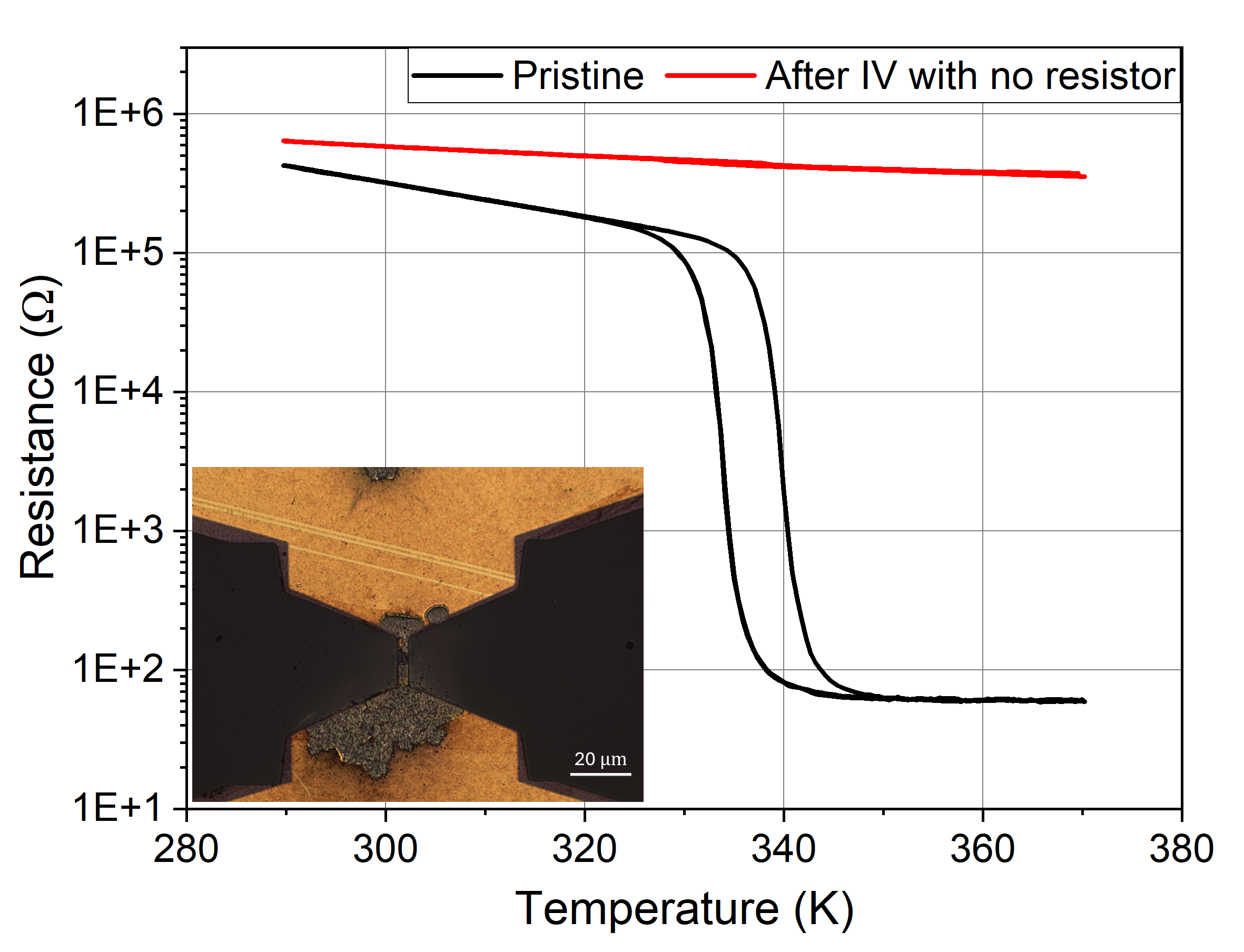}
         \caption{}
         \label{sup fig: RT no resistor}
     \end{subfigure}
     \hfill
     \begin{subfigure}[h]{0.48\linewidth}
         \centering
         \includegraphics[width=\textwidth]{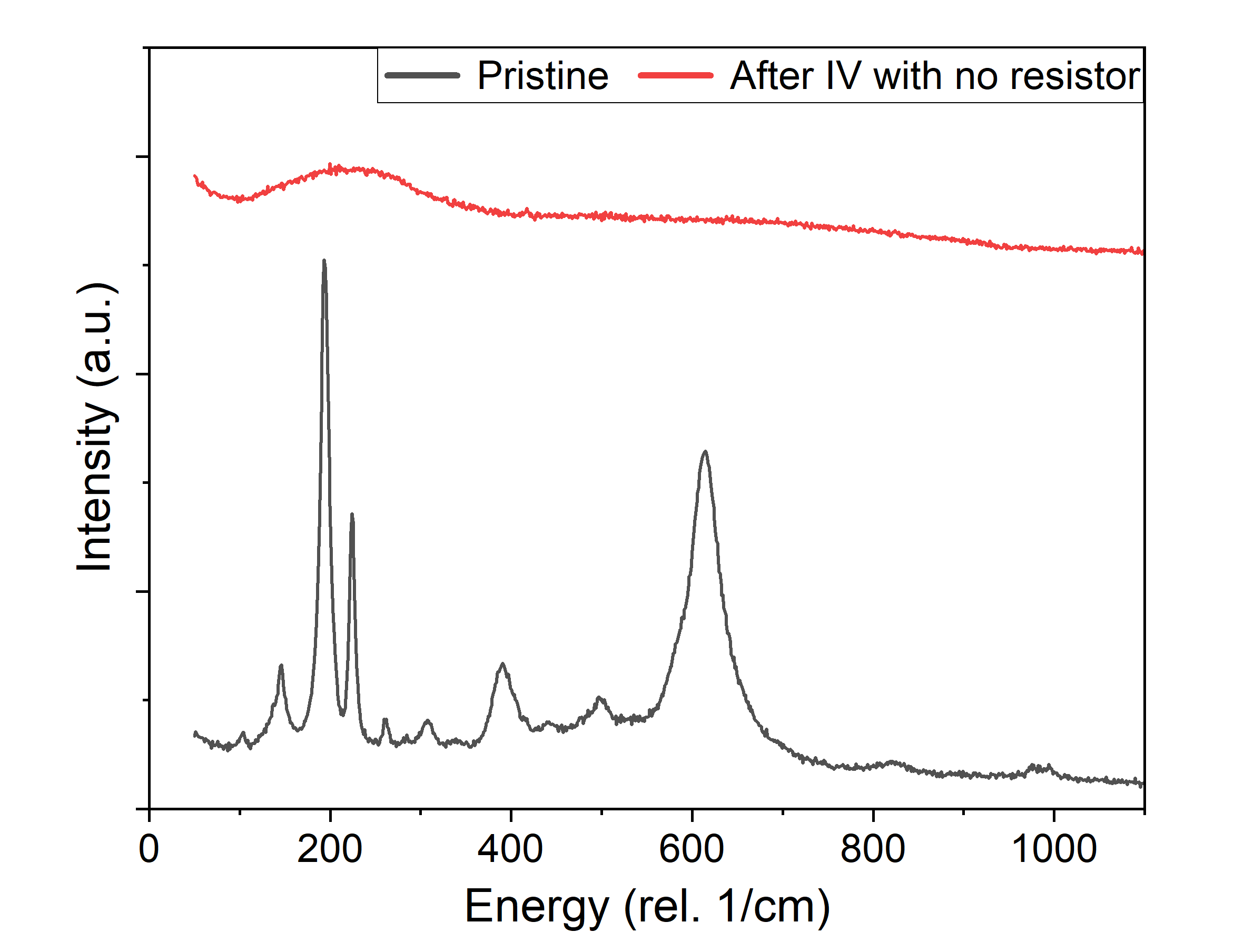}
         \caption{}
         \label{sup fig: raman no resistor}
     \end{subfigure}
     \hfill
    \caption{Damage induced by the current surge to a VO\textsubscript{2} device subjected to an IV measurement without protective resistor at T=300K: \textbf{(a)} RT curves of the device before and after the IV, showing no IMT features after the surge. Inset: Optical micrograph of the device after the IV, showing visible damage to the gap and the electrodes. \textbf{(b)} Raman spectra of the device before and after the IV, the crystalline features of the spectrum nearly disappeared after the IV, suggesting an almost complete amorphization of the device.} 
    \label{sup fig: damaged device no resistor}
\end{figure}

\newpage

\section{Role of the Internal Resistor in Device Protection}
\begin{figure}[!htbp]
     \centering
     \includegraphics[width=0.7\linewidth]{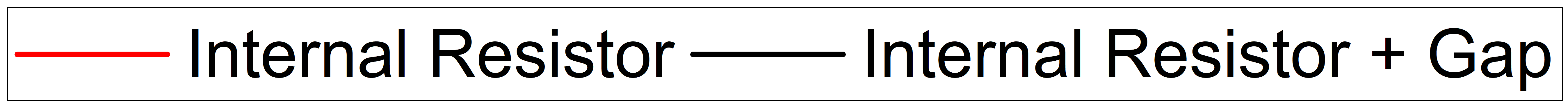}
     \begin{subfigure}[h]{0.48\textwidth}
         \centering
         \includegraphics[width=\textwidth]{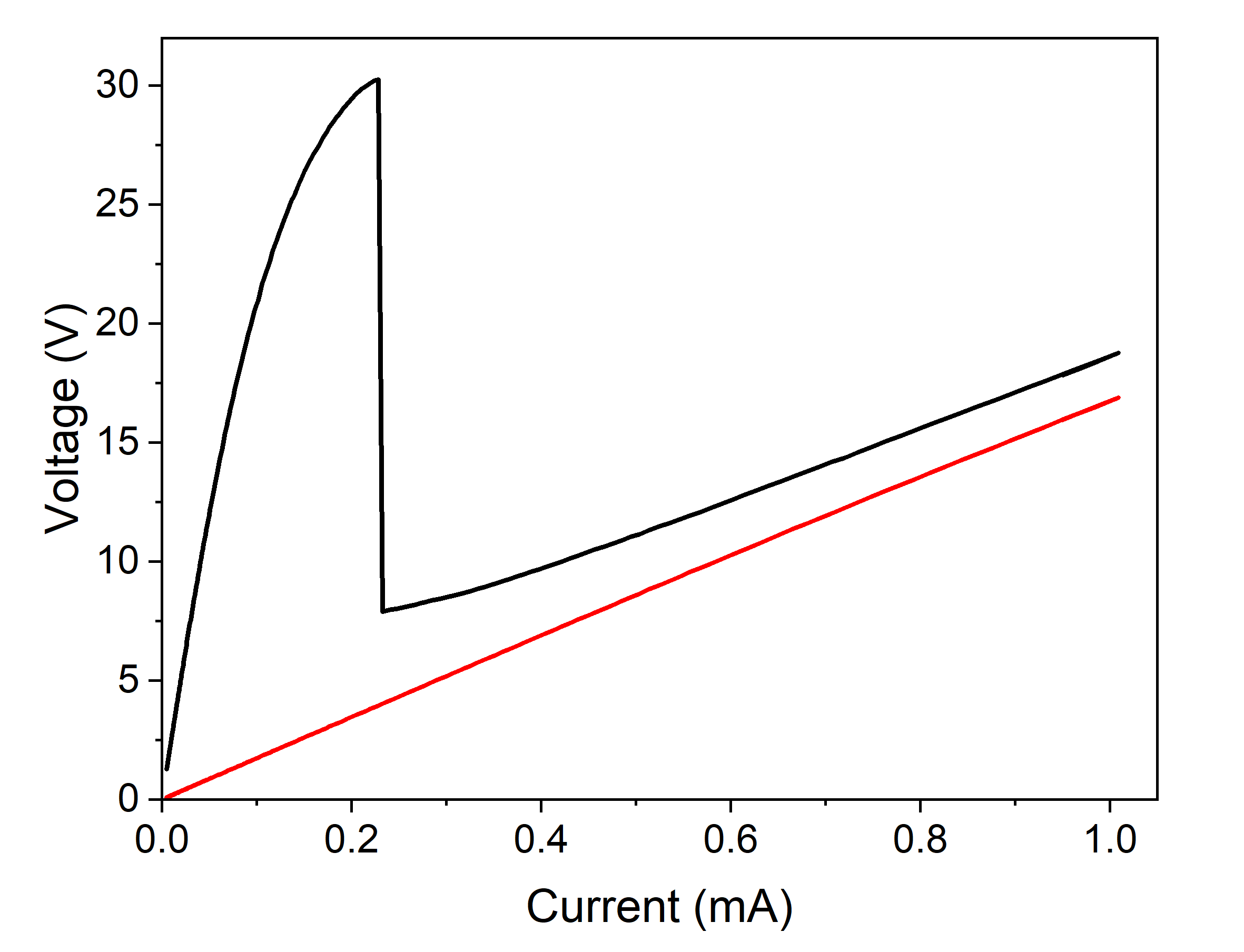}
         \caption{}
         \label{sup fig: Internal resistor IV graph}
     \end{subfigure}
     \hfill
     \begin{subfigure}[h]{0.48\textwidth}
         \centering
         \includegraphics[width=\textwidth]{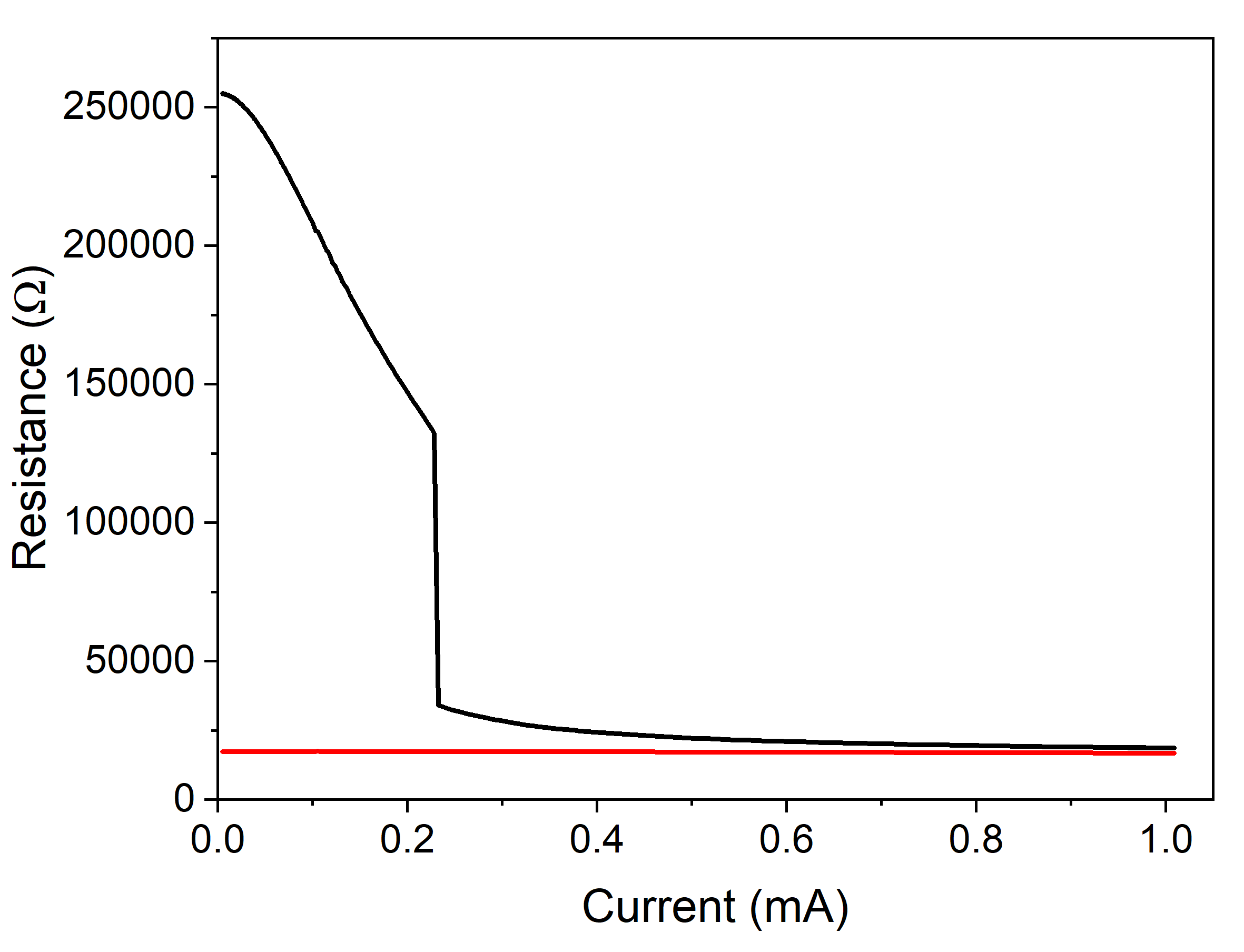}
         \caption{}
         \label{sup fig: Internal resistor IR graph.png}
     \end{subfigure}
     \hfill
    \caption{\textbf{(a)} Current vs voltage and \textbf{(b)} current vs resistance curves of the Internal resistor and the complete circuit of a VO\textsubscript{2} device at T=300K.}
    \label{sup fig: Internal Resistor curves}
\end{figure}

The use of an internal series resistor is only viable if the oxide resistor exhibits nearly ideal, bias-independent behavior, as discussed in the main text. This is demonstrated in Figure \ref{sup fig: Internal Resistor curves}. Although the complete circuit, comprising both the gap and the internal series resistor displays a pronounced switching event, the internal resistor measured independently retains an approximately linear and bias-independent IV curve. Nevertheless, for extraction of the IV curves of the gap, the internal resistor is measured independently and subtracted from the total circuits' measured voltage, accounting for non bias dependent behavior of the internal resistor. \vspace{6pt} 

Having established that the internal resistor behaves stable as a circuit element, we now examine its effectiveness in suppressing the current and power surges associated with resistive switching. As discussed in the main text, two types of current surge can occur following switching. The first type originates from the slow response of the SMU compared to the fast switching of the material, This type of surge can be at least partially suppressed by incorporating a resistor in series with the gap. The second arises from the discharge of parasitic capacitance in the measurement circuit through the newly formed metallic filament. This capacitive discharge can occur even for an ideal current source. In this case, the magnitude of the surge depends strongly on whether the series resistor is external or internal, i.e., lithographically defined as part of the device, since the two configurations involve different parasitic capacitances.\vspace{6pt}.

In this section, we estimate the damage induced by these current surges. We show that without protection, and even with an external series resistor, the surge can be sufficient to destroy the device. By contrast, incorporating an internal resistor can substantially reduce the surge-induced damage.\vspace{6pt}

The damage is estimated in terms of equivalent heating, defined as the temperature increase that would result if all the energy released during the current surge were converted into thermal energy within the gap volume. Since some of the input energy is dissipated through the substrate and electrodes, this estimate provides an upper bound on the ultimate temperature of the device. This estimate become more accurate when the surge timescale is shorter than the thermal dissipation timescale. Therfore, if the calculated final temperature exceeds the melting temperature of the device material, it is reasonable to conclude that the surge is sufficient to cause catastrophic device failure.

The increase in temperature can be written as:
\begin{equation}
    \Delta T = \frac{E_{surge}}{C_{P,v}\cdot\Omega_{gap}}
    \label{equivalent heating eq}
\end{equation}

Where E\textsubscript{surge} is the excess energy deposited in the gap resulting from the power surge, C\textsubscript{P,v} is the volumetric heat capacity of the material and $\Omega$\textsubscript{gap} is the volume of the defined gap.

The volumetric heat capacity of the metallic phase is $\sim3\times10^6 \frac{J}{K\cdot m^3}$ for both VO\textsubscript{2} \cite{kizuka_temperature_2015} and V\textsubscript{2}O\textsubscript{3} \cite{keer_thermal_1977}. The volume of the gap is $\Omega_{gap}=4\mu m\cdot 4\mu m\cdot 0.1\mu m=1.6\times 10^{-18} m^3$\vspace{6pt}

We now estimate E\textsubscript{surge}, beginning with an unprotected device. The excess energy deposited during the surge is determined by the duration over which the excess power heats the device before it is either dissipated to the surroundings, primarily through the substrate, or limited by the adjustment of the SMU output to the new low-resistance metallic state. The characteristic thermal timescale of our device is estimated to be on the order of tens of nanoseconds, \cite{del_valle_subthreshold_2019, hamaoui_thermophysical_2019} whereas the response time of the SMU is on the order of $\sim100\mu s$, Therefore, the surge energy is expected to be limited primarily by the thermal timescale.

The excess power is proportional to the resistance ratio between the insulating and the metallic phases. as shown below:\vspace{6pt}

Right before the switching, the switching power P\textsubscript{sw} is given by:
\begin{equation}
    P_{sw} = \frac{V_{sw}^2}{R_i}
    \label{switching power eq}
\end{equation}
When V\textsubscript{sw} is the switching voltage and R\textsubscript{i} is the resistance of the insulating state of the material. Right after the switching, the resistance switched from R\textsubscript{i} to R\textsubscript{m}, the resistance of the metallic state. The SMU continues to apply V\textsubscript{sw} on the device since it is slow to adjust to the new resistance, which results in a power surge:
\begin{equation}
    P_{surge} = \frac{V_{sw}^2}{R_m}
\end{equation}
Plugging V\textsubscript{sw} from equation \eqref{switching power eq}:
\begin{equation}
    P_{surge} = P_{sw}\frac{R_i}{R_m}
\end{equation}

since the ratio between the resistance of the insulating phase and the metallic phase is $\sim10^3$ for VO\textsubscript{2} and $\sim10^5$ for V\textsubscript{2}O\textsubscript{3} the resulting power surge is extremely large.

In our VO\textsubscript{2} devices, the switching power at 300K was $\sim6mW$, Substituting these values into \eqref{equivalent heating eq} we get:

\begin{equation}
    \Delta T = \frac{E_{surge}}{C_{P,v}\cdot\Omega_{gap}}=\frac{P_{surge}\tau_{thermal}}{C_{P,v}\cdot\Omega_{gap}}=\frac{P_{sw}\frac{R_i}{R_m}\tau_{thermal}}{C_{P,v}\cdot\Omega_{gap}}\approx\frac{6mW\cdot10^3\cdot10ns}{3\times10^6 \frac{J}{K\cdot m^3}\cdot1.6\times 10^{-18} m^3}\sim10^4K
    \label{equivalent heating no resitor eq}
\end{equation}

The device is expected to melt before reaching this estimated temperature, consistent with the experimental observations presented in Supplementary Section S2. In V\textsubscript{2}O\textsubscript{3}, where the resistance ratio between the insulating and metallic phases is even higher, an unprotected device is likewise expected to undergo catastrophic failure.\vspace{6pt}

When a resistor in incorporated in series with the gap, the current surge can be substantially reduced, as the surge now depend on the resistance ratio between the insulating phase resistance and the resistor resistance:

Right before the switching, the switching power P\textsubscript{sw} is given by:
\begin{equation}
    P_{sw} = \frac{V_{sw}^2}{R_i+R_{resistor}}
    \label{switching power w resistor eq}
\end{equation}

Right after the switching We have:
\begin{equation}
    P_{surge} = \frac{V_{sw}^2}{R_m+R_{resistor}}\approx\frac{{V_{sw}^2}}{R_{resistor}}=P_{sw}\frac{R_i+R_{resistor}}{R_{resistor}}
    \label{surge power w resistor eq}
\end{equation}

The resistance of the gap can be neglected after the switching as it is orders of magnitude lower than the resistance of the resistor.
In an internal resistor, The difference in resistance from the gap stems entirely from the difference in their geometry. The devices in this work are made from a 4$\mu$m-by-4$\mu$m gap and a 450$\mu$m-wide by 40$\mu$m-long internal resistor. Since their thickness and resistivity are the same. Their resistance ratio amounts to:
\begin{equation}
\frac{R_i}{R_{resistor}} = \frac{\rho_{VO_x}\frac{L_{gap}}{W_{gap}}}{\rho_{VO_x}\frac{L_{resistor}}{R_{resistor}}} = \frac{\frac{4\mu m}{4\mu m}}{\frac{40\mu m}{450\mu m}} \approx 11
\end{equation}
This resistance ratio can also be extracted from Figure \ref{sup fig: Internal resistor IR graph.png}. Entering this value into equation \eqref{surge power w resistor eq}:
\begin{equation}
    P_{surge} = P_{sw}\frac{R_i+R_{resistor}}{R_{resistor}} \approx 12P_{sw} 
\end{equation}
The power surge is substantially reduced compared with the unprotected case, where the surge is at least two orders of magnitude larger. As a result, catastrophic device failure is avoided. Nevertheless, the remaining surge can still modify the device, as shown in the main text. \vspace{6pt}

The importance of using an internal resistor becomes clearer when considering the discharge of parasitic capacitance. Even if the SMU-related current surge is mitigated by placing a resistor in series with the device, the energy stored in the parasitic capacitance of the measurement circuit can still be released through the metallic filament after switching. This excess energy is given by
\begin{equation}
    E_{surge} = \frac{1}{2}C_{parasitic}V_{sw}^2 
\end{equation}

For an external series resistor, the relevant capacitance can be estimated from the parasitic capacitance of the coaxial cables, which is typically on the order of $\sim100pF/m$. Here, we use 50pF as an estimate for the circuit capacitance. The corresponding equivalent heating is:
\begin{equation}
    \Delta T = \frac{E_{surge}}{C_{P,v}\cdot\Omega_{gap}} = \frac{\frac{1}{2}C_{parasitic}V_{sw}^2}{C_{P,v}\cdot\Omega_{gap}}=\frac{\frac{1}{2}\cdot50pF\cdot(50V)^2}{3\times10^6 \frac{J}{K\cdot m^3}\cdot1.6\times 10^{-18} m^3}\sim10^4K 
\end{equation}
This estimated temperature increase is more than sufficient to melt the device. The characteristic discharge time is $\tau=R_mC_{parasitic}\approx100\Omega\cdot50pF=5ns$. which is on the order of the thermal timescale of the device. A substantial fraction of the stored capacitive energy is deposited as heat before it can dissipate to the surroundings.

When the resistor is internal, the parasitic capacitance directly in parallel with device is reduced by orders of magnitude. In this configuration, the dominant capacitive element is the charge accumulated across the gap while it is in the insulating state. The gap can therefore be approximated as a parallel-plate capacitor:
\begin{equation}
    C_{gap}=\frac{\varepsilon_r\varepsilon_0W_{gap}t}{d}
\end{equation}
Where \textit{$\varepsilon_r$} is the static dielectric constant of the material in the gap (i.e. VO\textsubscript{2} or V\textsubscript{2}O\textsubscript{3}), \textit{W\textsubscript{gap}} is the width of the electrode, \textit{t} is the film thickness and \textit{d} is the gap length.

The static dielectric constants of the insulating phase of VO\textsubscript{2} and V\textsubscript{2}O\textsubscript{3} are $\sim35$ \cite{yang_dielectric_2010} and $\sim5000$ \cite{ma_dielectric_2023}, respectively. Using $W_{gap}=4\mu m,$, $t\sim100nm$ and $d=4\mu m$,
the estimated gap capacitances are:
\begin{align*}
    C_{gap,VO_2}=0.03fF  \\
    C_{gap,V_2O_3}=4fF
\end{align*}

For a V\textsubscript{2}O\textsubscript{3} device with an internal resistor, the corresponding equivalent heating is:
\begin{equation}
    \Delta T = \frac{E_{surge}}{C_{P,v}\cdot\Omega_{gap}} = \frac{\frac{1}{2}C_{parasitic}V_{sw}^2}{C_{P,v}\cdot\Omega_{gap}}=\frac{\frac{1}{2}\cdot4fF\cdot(50V)^2}{3\times10^6 \frac{J}{K\cdot m^3}\cdot1.6\times 10^{-18} m^3}\sim1K 
\end{equation}

In VO\textsubscript{2}, the gap capacitance is even smaller, and the corresponding equivalent heating is therefore also lower. The characteristic discharge time for the V\textsubscript{2}O\textsubscript{3} estimate is $\tau=R_mC_{parasitic}\approx100\Omega\cdot4fF=0.4ps$, much shorter than the thermal timescale of the device. Thus, the stored capacitive energy is expected to be deposited rapidly in the device, but its total magnitude is too small to cause catastrophic damage, in contrast to the case of an external resistor. Local heating within the filamentary region may still be higher than the spatially averaged estimate and could potentially contribute to local defect formation, nevertheless, global device damage is avoided.

\newpage
 
\section{Changes to the RT Curve of the Device after the Observed Memory Effect}
As discussed in the main text, the observation of the memory effect was not accompanied by a significant change in the shape of the RT curve. However, the measured RT response may underestimate the effect of the IV sweep on the insulating-phase resistance within the active switching region. In the original device geometry, the RT measurement probes the gap area in parallel with larger surrounding regions that may remain unaffected by the IV sweep. As a result, the measured resistance represents a weighted average that is dominated by these larger peripheral regions.

\begin{figure}[!htbp]
\centering
\includegraphics[width=0.8\textwidth]{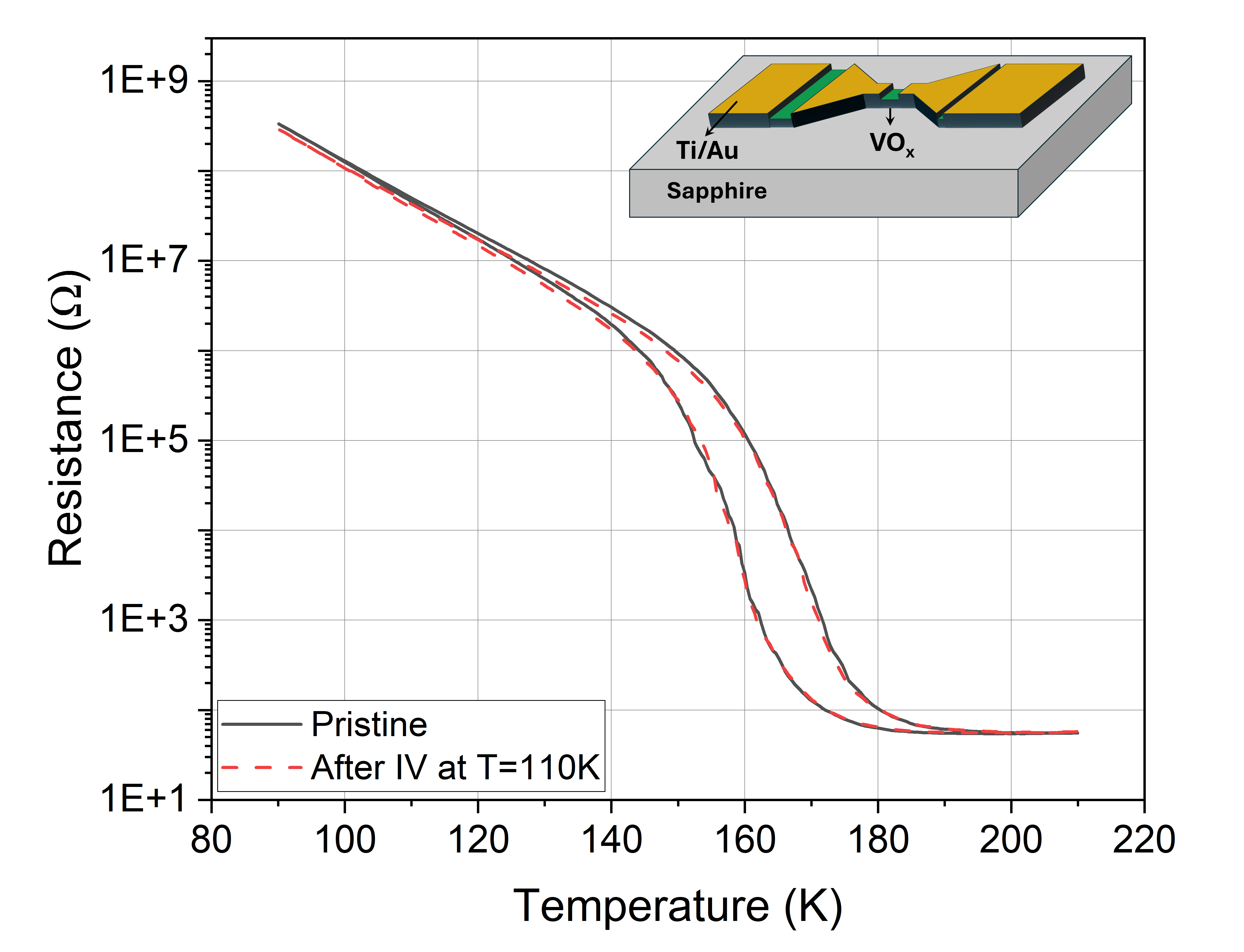}
\caption{Comparison of the RT curves before an after an etched V\textsubscript{2}O\textsubscript{3} device was subjected to an IV at T=110K. Inset: model of an etched device}
\label{sup fig: RT before after etched device}
\end{figure}

To better resolve possible IV-induced changes within the electrode gap, we fabricated a similar device and used a Reactive Ion Etching Process in a Plasma-Therm 790 to remove all material except the region directly between the electrodes (Device model shown in Figure \ref{sup fig: RT before after etched device} inset). This etched geometry also better represents the architecture of scaled devices. The RT curve was then measured before and after an IV sweep. As shown in Figure \ref{sup fig: RT before after etched device}, the two curves remain very similar, suggesting that any IV-induced modification is highly localized and remains largely masked by the unaffected portions of the gap.

\newpage

\section{Stochasticity in the Switching in the Phase Coexistence Regime and its Effect on the Observed Memory in V\textsubscript{2}O\textsubscript{3}}

During measurements performed within the phase coexistence regime, it is important to account for the inherently stochastic and temperature-sensitive nature of the switching process. In the phase-coexistence regime, switching can be triggered by the transformation of a single critical domain that completes a percolating metallic path across the device. Consequently, small changes in either the spatial distribution of metallic and insulating domains or the measurement temperature can produce large variations in the initial device resistance, and therefore in the observed switching voltage. In this section, we examine whether the observed memory effect remains distinguishable from this intrinsic stochasticity, and how its magnitude varies across repeated IV sweeps and thermal-cycling steps.\vspace{6pt}

\begin{figure}[!htbp]
     \centering
     \includegraphics[width=\linewidth]{Figures/multiple_sweeps_legend.png}
     \begin{subfigure}[h]{0.48\textwidth}
         \centering
         \includegraphics[width=\textwidth]{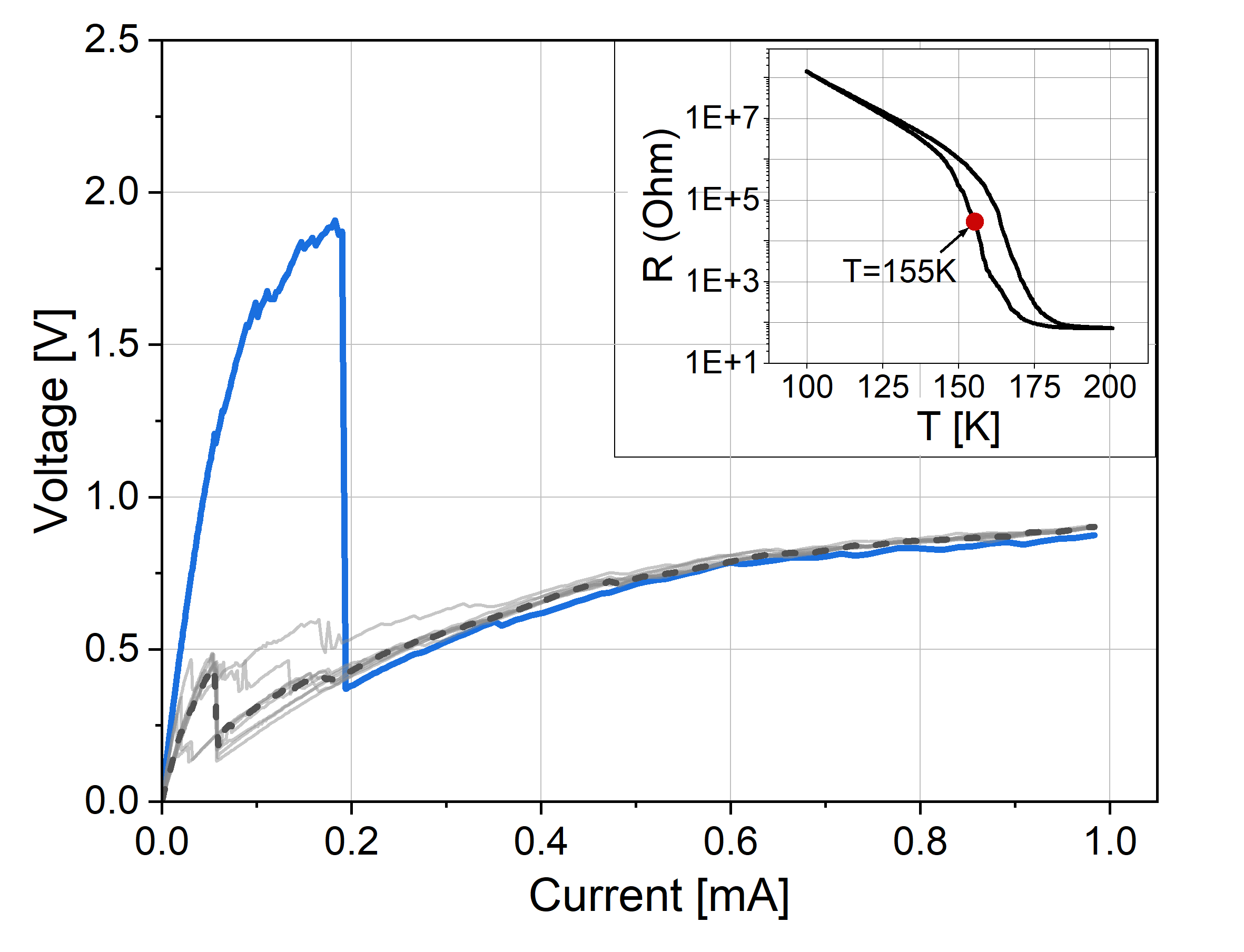}
         \caption{}
         \label{fig: memory high T V2O3}
     \end{subfigure}
     \hfill
     \begin{subfigure}[h]{0.48\textwidth}
         \centering
         \includegraphics[width=\textwidth]{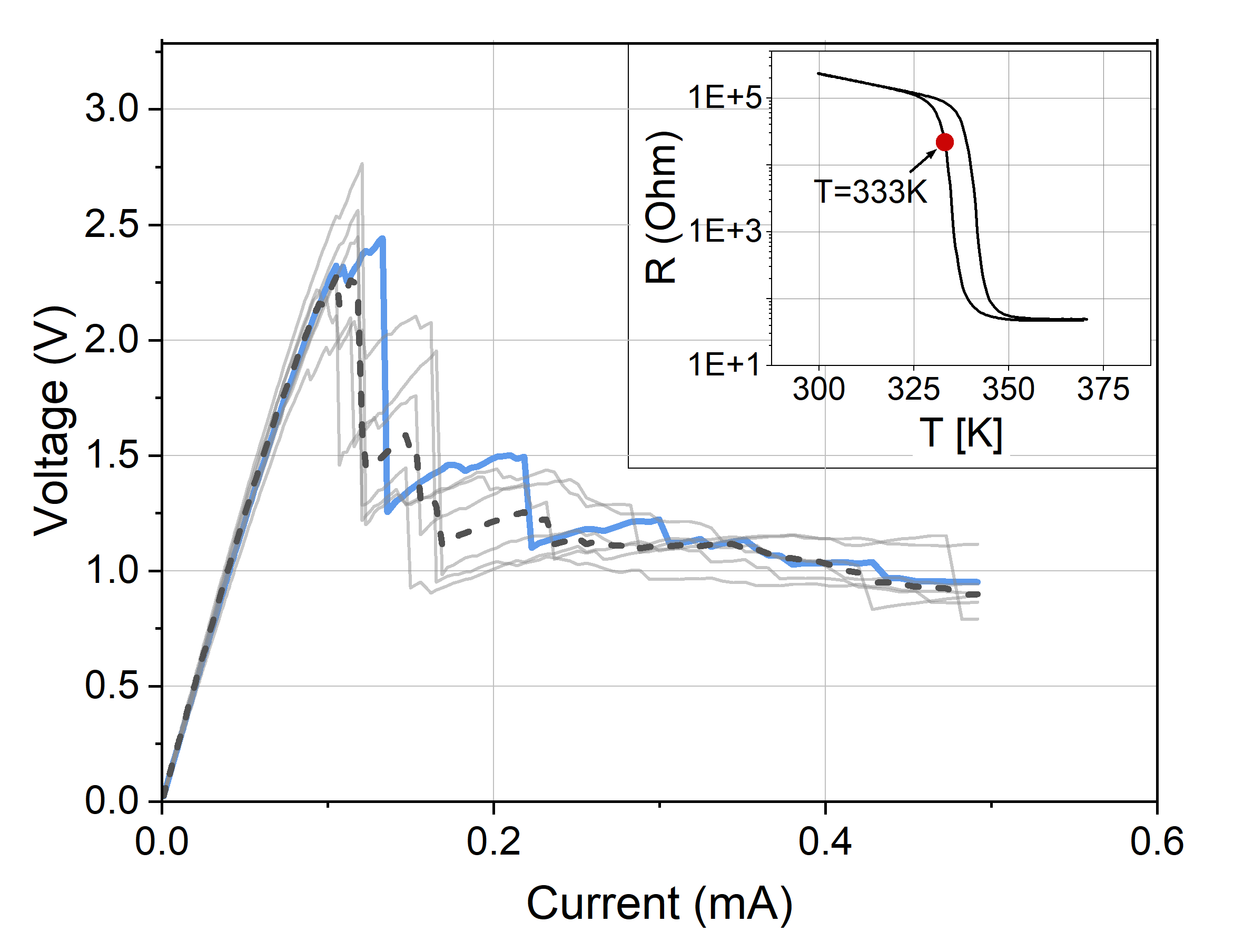}
         \caption{}
         \label{sup fig: memory high T VO2}
     \end{subfigure}
     \hfill
    \caption{Comparison between the first IV sweep and subsequent sweeps in \textbf{(a)} V\textsubscript{2}O\textsubscript{3}, where a memory effect is observed, and in \textbf{(b)} VO\textsubscript{2}, where it is not. Inset: The location of the measurement temperature on the R(T) curve of the device.}
    \label{sup fig: memory high T}
\end{figure}

Figure \ref{sup fig: memory high T} shows the results of repeated IV measurements performed inside the hysteresis loop in VO\textsubscript{2} and V\textsubscript{2}O\textsubscript{3}. In both materials, the shape of the IV curve and the switching voltage fluctuates between consecutive sweeps. In V\textsubscript{2}O\textsubscript{3}, but not in VO\textsubscript{2}, the variation among individual IV sweeps
is smaller than the systematic difference between the first IV sweep and subsequent sweeps.

\begin{figure}[h]
\centering
\includegraphics[width=0.6\textwidth]{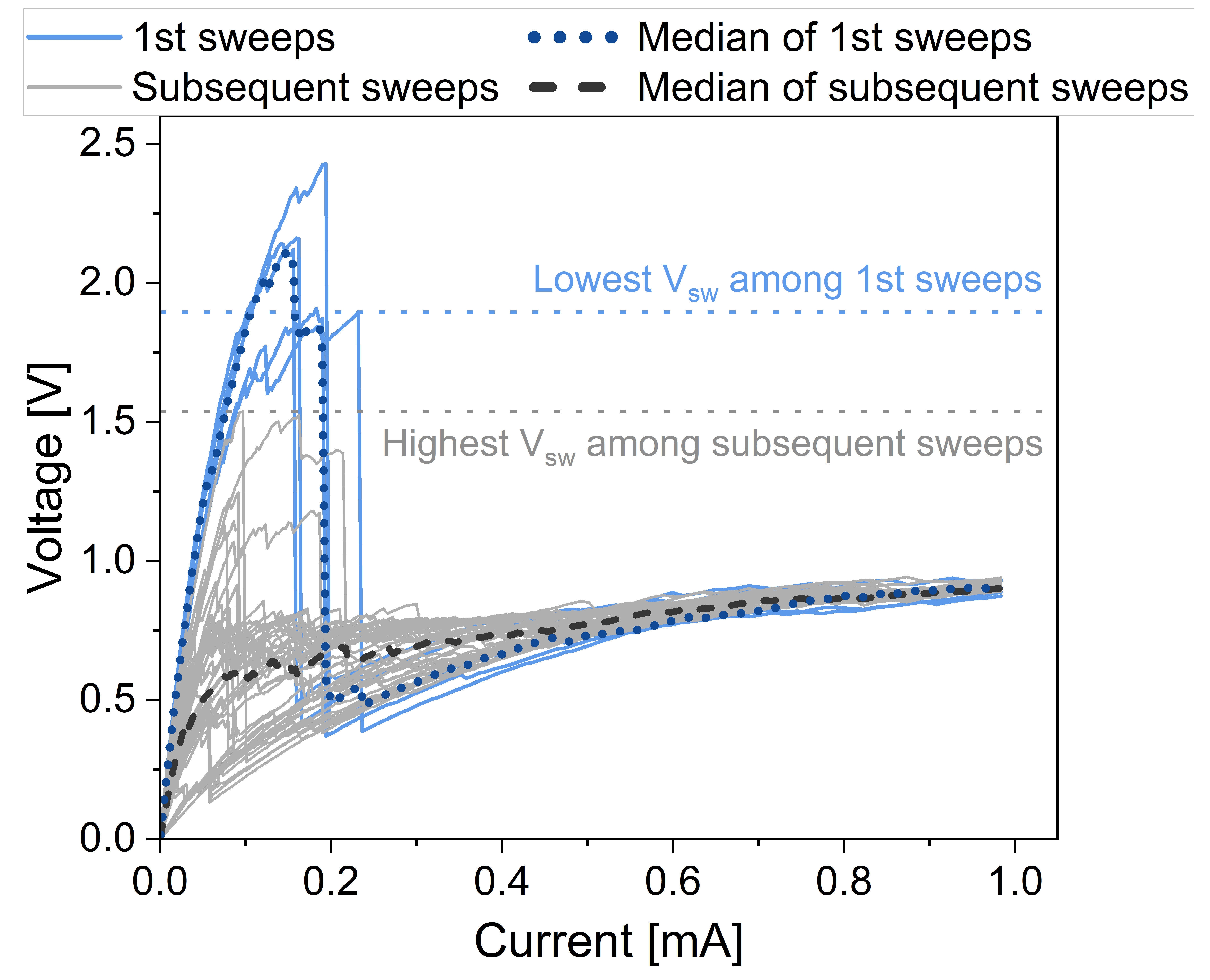}
\caption{The IV curves of 50 consecutive sweeps, with every 10 sweeps separated by a thermal cycling step. The blue curves are the first IV sweeps after each thermal cycling step. The measurement were performed on a V\textsubscript{2}O\textsubscript{3} device in the phase coexistence region (T=155K). }
\label{sup fig: memory inside hystersis all curves}
\end{figure}

To further test the robustness of this effect under more variable conditions, we performed repeated IV measurements separated by thermal-cycling steps. In addition to resetting the domain configuration, thermal cycling may introduce additional stochasticity through cycle-to-cycle variations in domain nucleation and small temperature fluctuations during repeated heating and cooling. In each round, the device was subjected to 10 consecutive IV measurements, followed by thermal cycling to erase the domain configuration. This procedure was repeated five times, yielding a total of 50 IV measurements with four thermal-cycling steps between successive rounds. The results are shown in Figure \ref{sup fig: memory inside hystersis all curves}. Although substantial stochasticity is observed among individual IV sweeps, including between the first IV sweeps of different rounds, the difference between the first IV sweep and subsequent sweeps remains robust. This behavior suggests that the first IV induces a non-volatile rearrangement of metallic and insulating domains, which lowers the switching voltage in subsequent cycles. 

\begin{figure}[!htbp]
     \centering
     \begin{subfigure}[h]{0.48\textwidth}
         \centering
         \includegraphics[width=\textwidth]{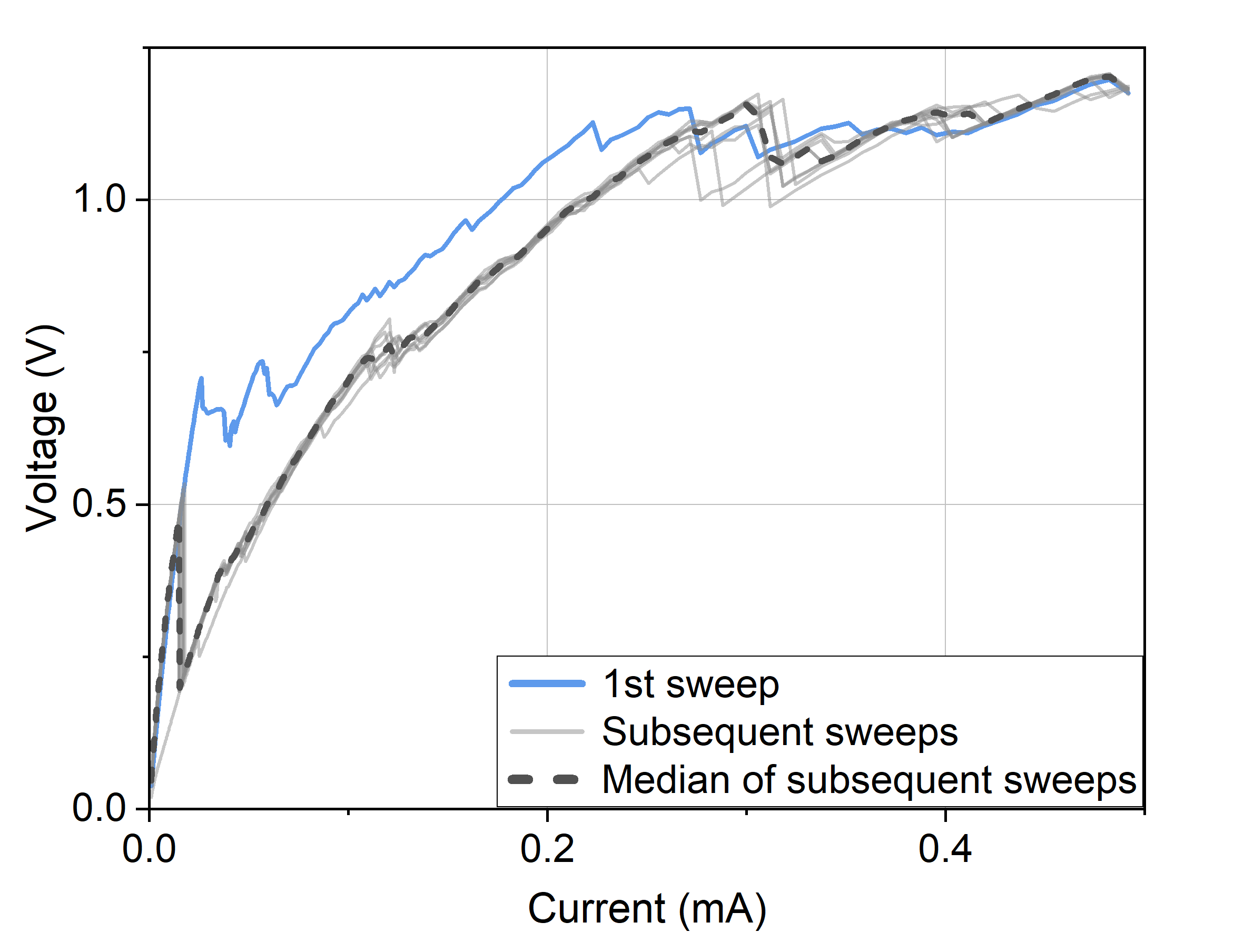}
         \caption{}
         \label{sup fig: memory inside hystersis weak memory}
     \end{subfigure}
     \hfill
     \begin{subfigure}[h]{0.48\textwidth}
         \centering
         \includegraphics[width=\textwidth]{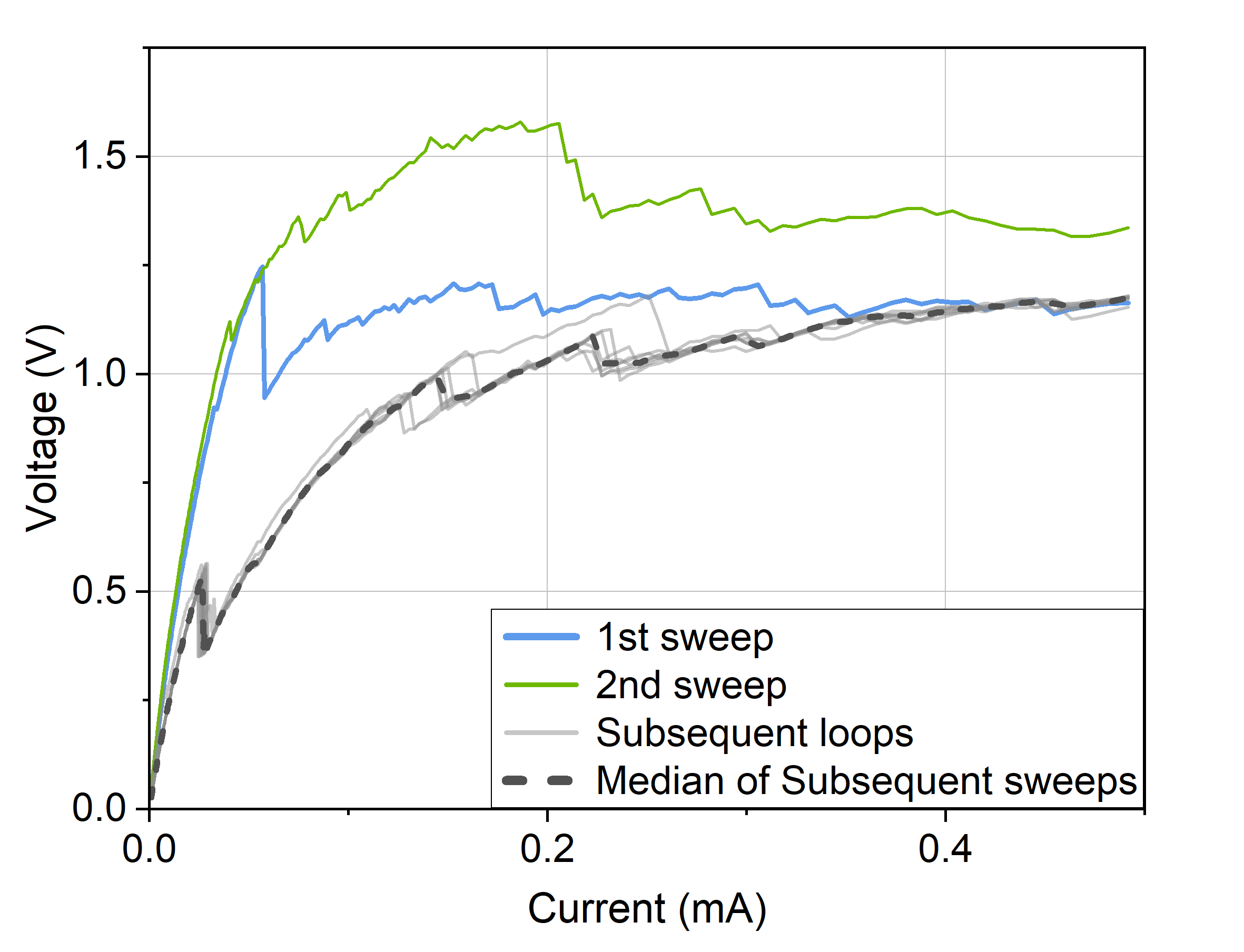}
         \caption{}
         \label{sup fig: memory inside hystersis after 2nd sweep}
     \end{subfigure}
     \hfill
    \caption{Effects of switching stochasticity on the phase-coexistence memory effect in V\textsubscript{2}O\textsubscript{3} devices. \textbf{(a)} The first IV sweep initially differs from subsequent sweeps, but the curves converge at higher currents, resulting in a similar measured switching voltage. \textbf{(b)} Both the first and second IV sweeps differ from subsequent sweeps.}
    \label{fig: memory inside hystersis weak and 2nd loop}
\end{figure}

Nevertheless, in some devices the stochasticity played a more prominent role. We observed cycles in which the distinction between the first sweep and subsequent sweeps was more subtle (Figure \ref{sup fig: memory inside hystersis weak memory}). In these cases, the measured switching voltages can be similar because the stochastic variation among sweeps partially masks the memory effect. We also observed cases in which the effect emerged only after more than one cycle (Figure \ref{sup fig: memory inside hystersis after 2nd sweep}). While the domain configuration produced by the first switching event typically lowers the voltage required to re-create the filament, stochastic domain redistribution can occasionally produce a configuration in which the switching voltage remains high.

\newpage

\section{SEM Image of a Pristine Device}
\begin{figure}[!htbp]
\centering
\includegraphics[width=0.55\textwidth]{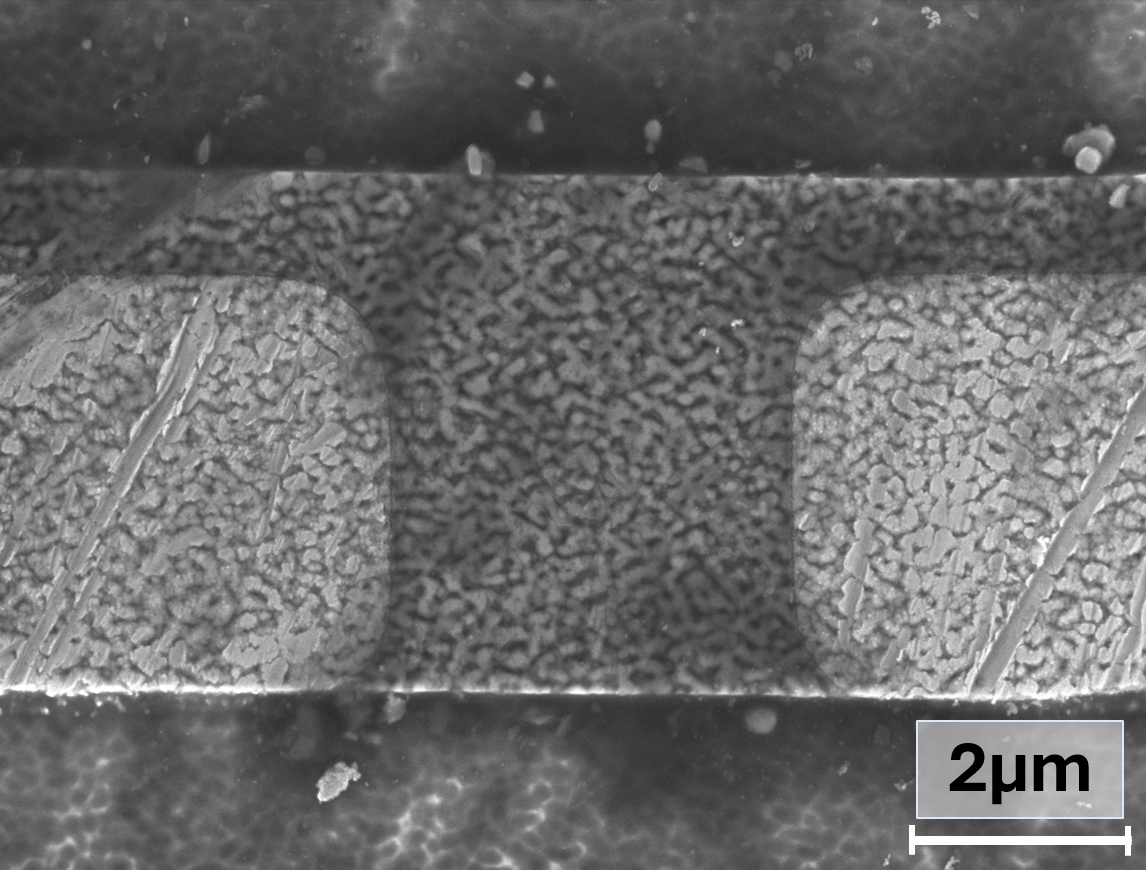}
\caption{SEM micrograph of a pristine VO\textsubscript{2} device.}
\label{sup fig: SEM Pristine Device}
\end{figure}

Figure \ref{sup fig: SEM Pristine Device} shows a SEM micrograph of a pristine VO\textsubscript{2} device. The observed morphology is similar to the device that went through a cumulative writing procedure (Figure 7c in the main text), indicating that this writing procedure produces no structural damage in the device. 

\section{Effect of Mild Annealing on the Defect-related Memory in VO\textsubscript{x} Devices}

\begin{figure}[!htbp]
     \centering
     \includegraphics[width=0.7\textwidth]{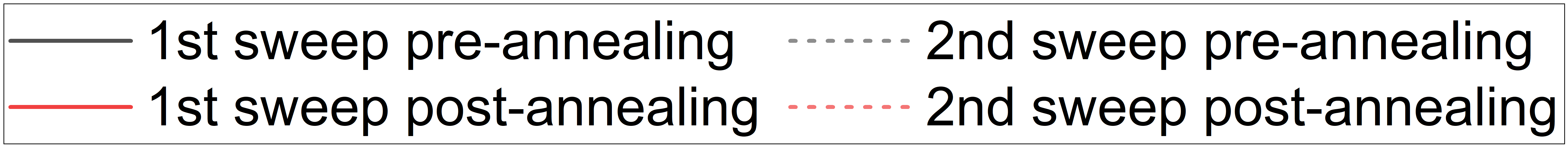}
     \begin{subfigure}[h]{0.48\textwidth}
         \centering
         \includegraphics[width=\textwidth]{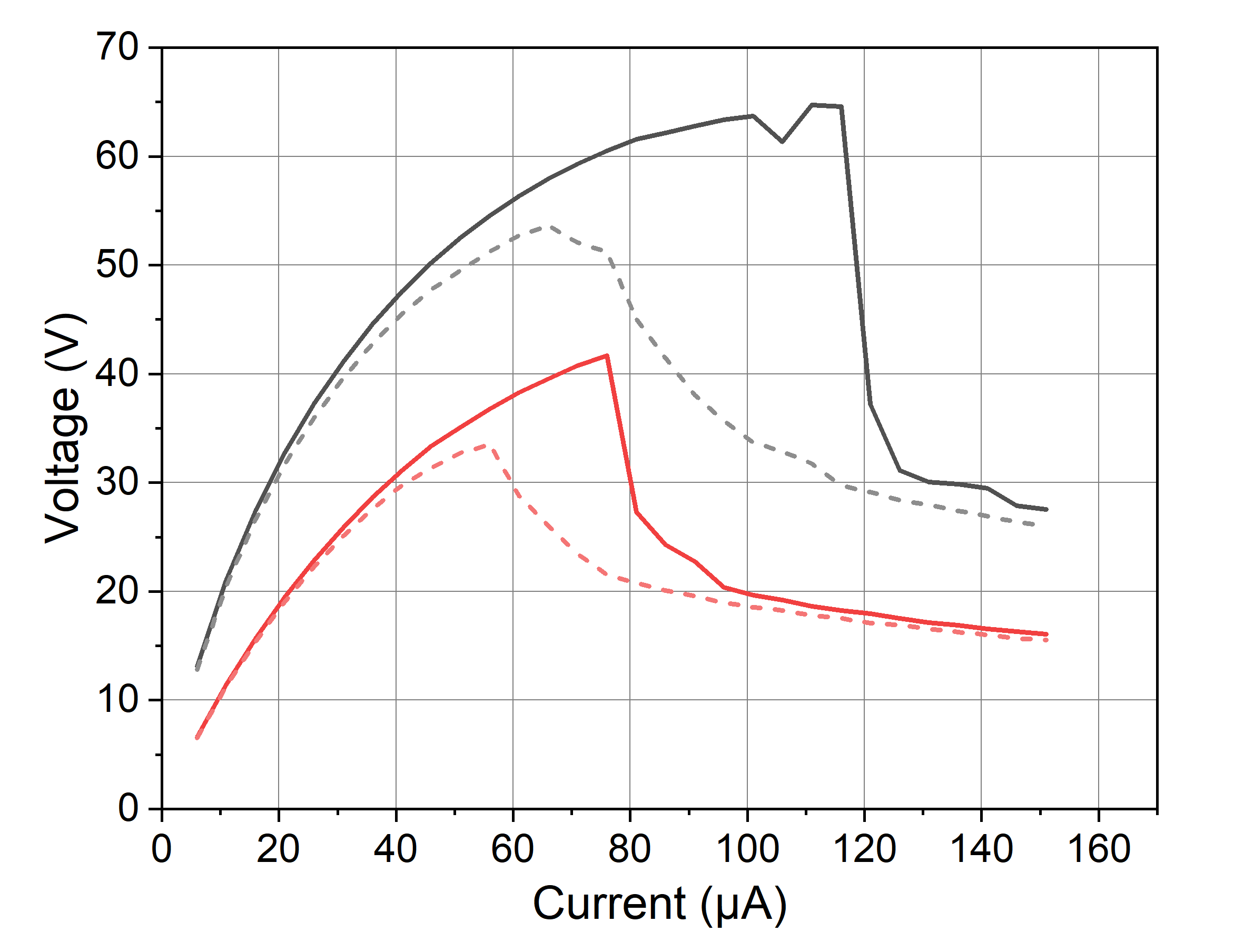}
         \caption{}
         \label{sup fig: annealing v203}
     \end{subfigure}
     \hfill
     \begin{subfigure}[h]{0.48\textwidth}
         \centering
         \includegraphics[width=\textwidth]{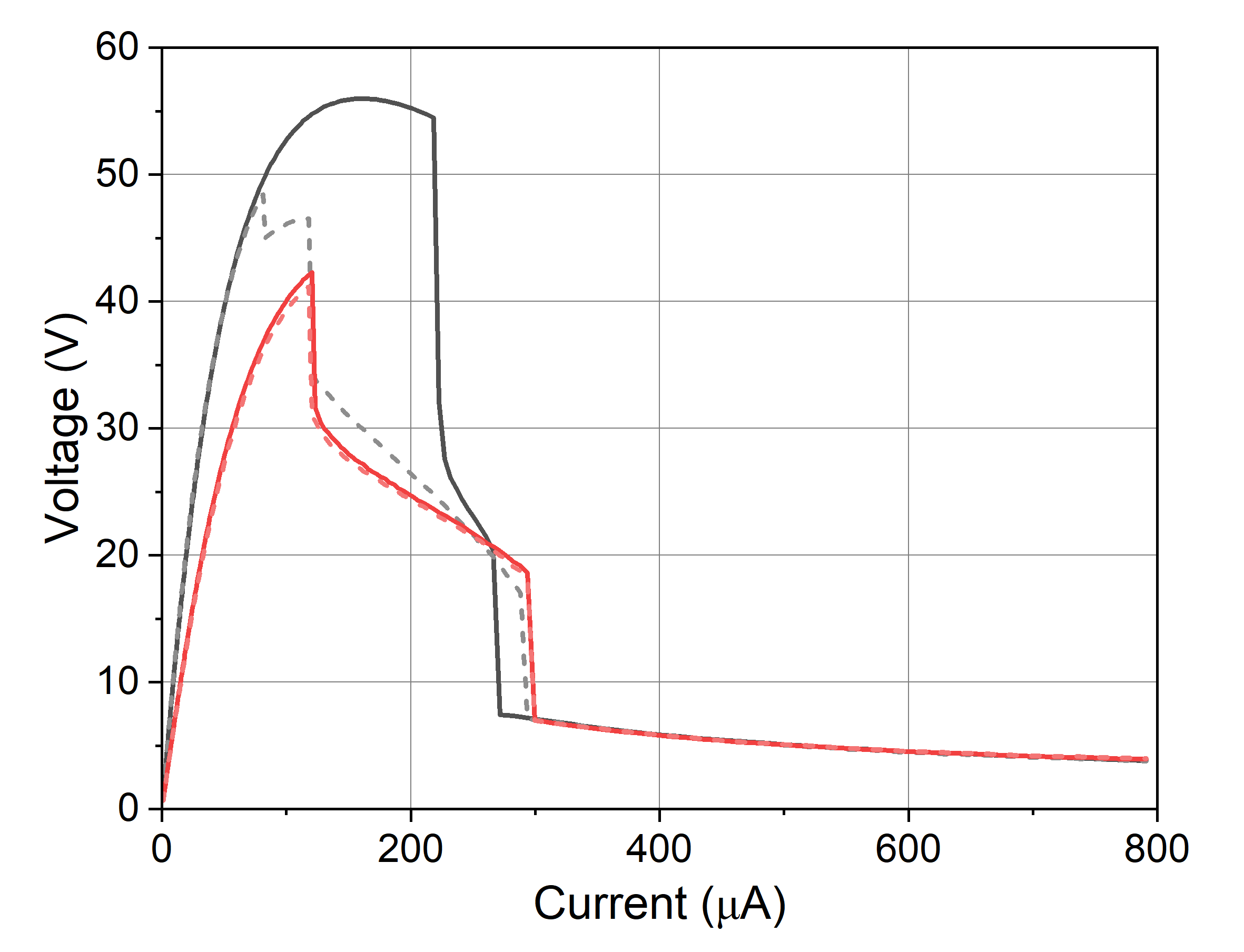}
         \caption{}
         \label{sup fig: annealing vo2}
     \end{subfigure}
     \hfill
     \begin{subfigure}[h]{0.48\textwidth}
         \centering
         \includegraphics[width=\textwidth]{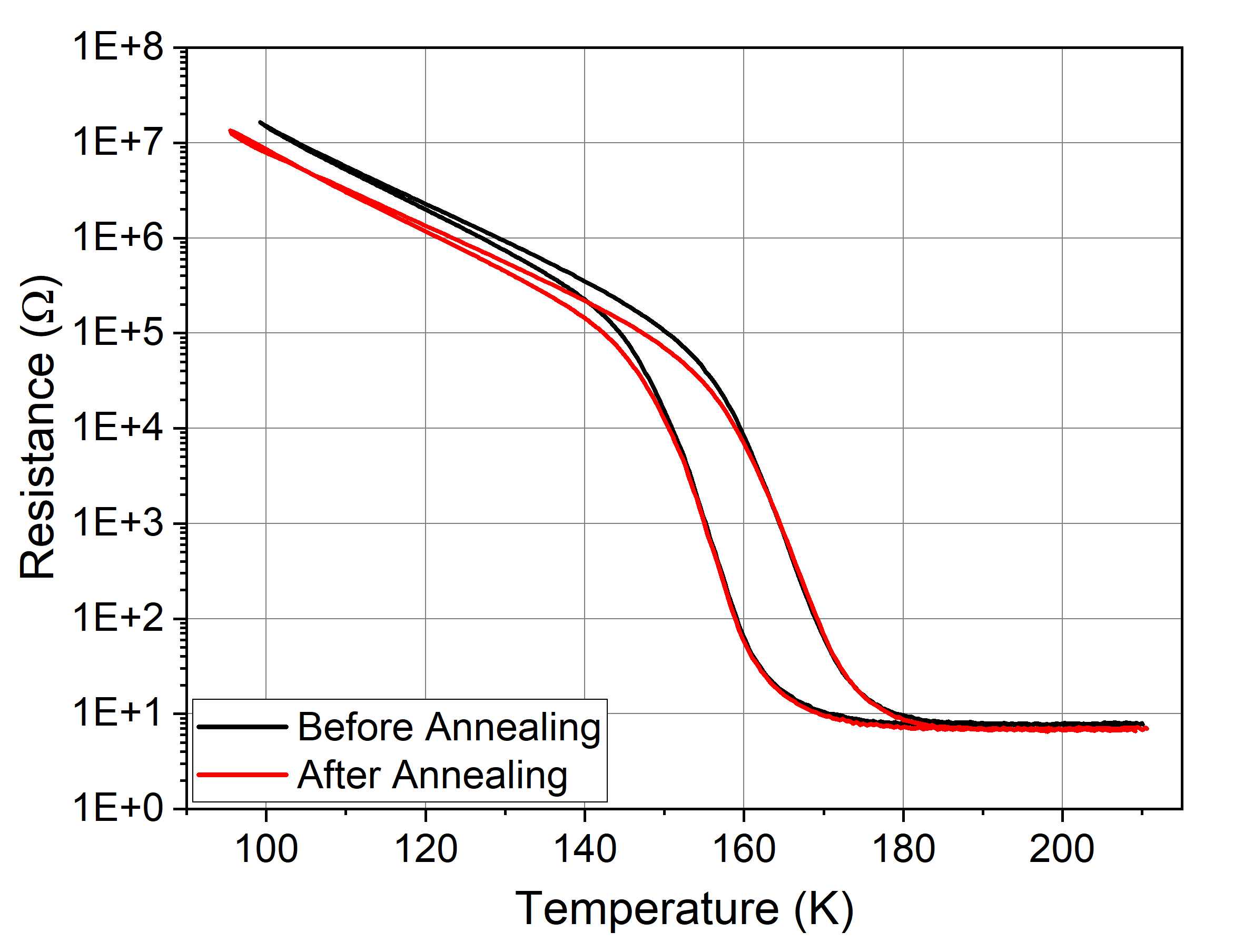}
         \caption{}
         \label{sup fig: annealing v203 RT}
     \end{subfigure}
     \hfill
     \begin{subfigure}[h]{0.48\textwidth}
         \centering
         \includegraphics[width=\textwidth]{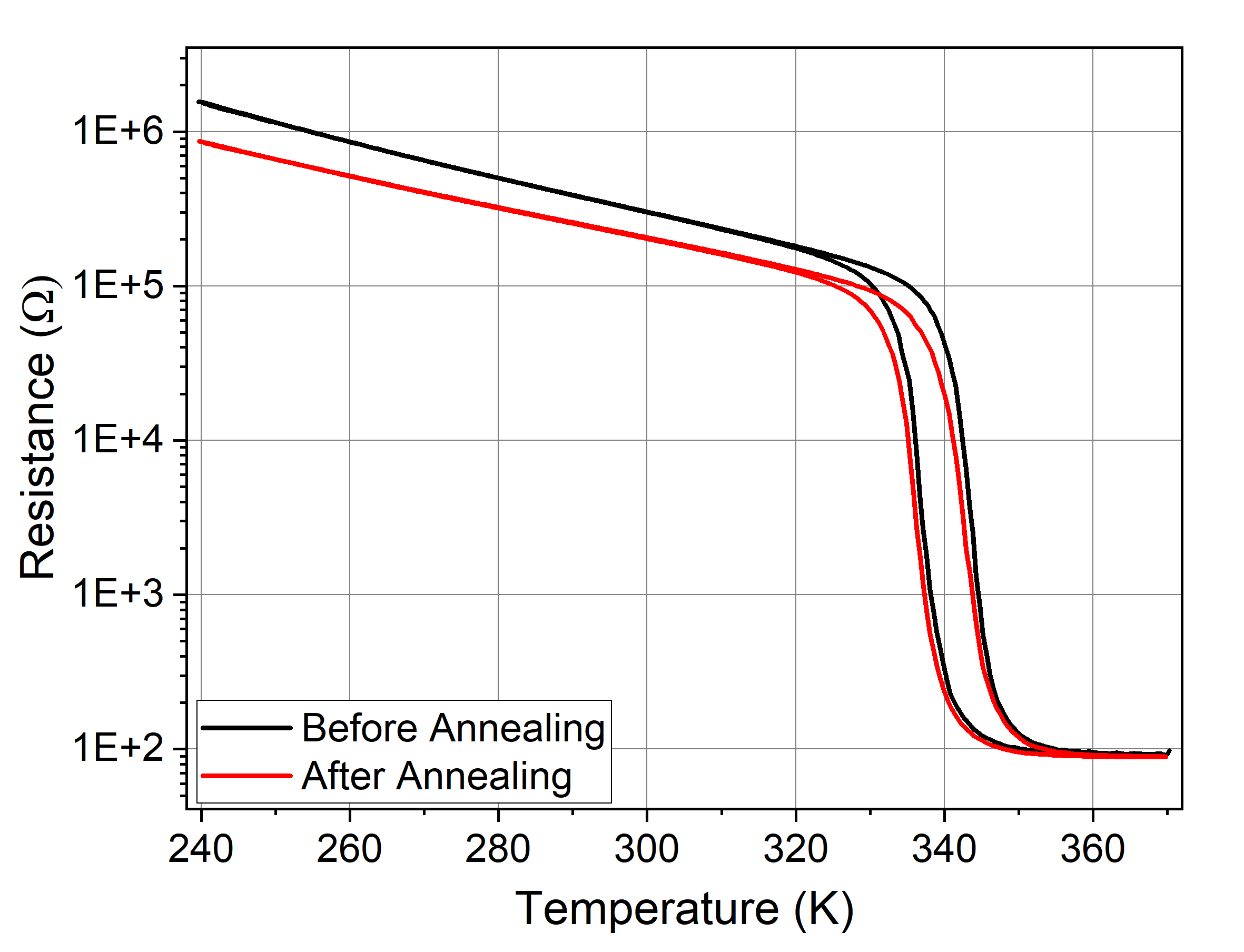}
         \caption{}
         \label{sup fig: annealing vo2 RT}
     \end{subfigure}
    \caption{Effects of annealing treatment at 400K for 10 hours in vacuum on the defect-related memory effect in VO\textsubscript{x} devices. \textbf{(a)} Two consecutive IV sweeps of a V\textsubscript{2}O\textsubscript{3} device before and after the annealing treatment. The measurement was taken at the fully-insulating state at T=115K. \textbf{(b)} Two consecutive IV sweeps of a VO\textsubscript{2} device before and after the annealing treatment. The measurement was taken at the fully-insulating state at T=250K. \textbf{(c-d)} The RT curves of the respecting films before and after the annealing treatment.}
    \label{sup fig: annealing}
\end{figure}

Having identified the low-temperature memory effect as defect-related, we investigated its response to annealing treatments. Annealing treatments can both introduce defects into the film and enhance the mobility or redistribution of existing defects. We therefore performed a series of mild thermal treatments in vacuum on our devices. Treatments below 400K produced no measurable effect. The influence of annealing at 400K for 10h in vacuum on the switching behavior and RT curves of V\textsubscript{2}O\textsubscript{3} and VO\textsubscript{2} devices is shown in Figure \ref{sup fig: annealing}.

Both oxides exhibit qualitatively similar behavior following annealing. The treatment produces a mild decrease in the insulating-phase resistance, similar to the change observed after the cumulative writing procedure in VO\textsubscript{2} devices (figure 7a in main text). This similarity supports the interpretation that the writing process is defect-related. As a result of the reduced insulating-phase resistance, the switching voltage decreases substantially after annealing.

Interestingly, the memory effect reappears after annealing, particularly in V\textsubscript{2}O\textsubscript{3}. This re-emergence may indicate that annealing not only increases the defect concentration, but also redistributes the defects more uniformly throughout the device. Subsequent IV cycling can then relocalize defects along the filamentary path, producing a renewed memory effect. However, it remains unclear why this relocalization occurs so readily, given that the lower switching voltage after annealing should reduce the magnitude of the current surge.

Since the annealing treatments were performed in vacuum, they are expected to promote defect formation, most likely through oxygen vacancy generation [ref. 62-63 in main text]. It is therefore possible that annealing under a controlled oxygen environment could instead heal oxygen-related defects and erase, or at least reduce, the defect-related memory effect. Further research is needed to clarify the relation between annealing atmosphere, defect redistribution, and the resulting memory behavior.

\renewcommand{\refname}{Supplementary References}
\bibliography{references}